\documentclass[11pt,a4paper]{article}
\usepackage{authblk}

\usepackage{comment}
\usepackage[english]{babel}
\usepackage[T1]{fontenc}
\usepackage[utf8]{inputenc}
\usepackage{soul}
\usepackage{color}
\usepackage{amsmath,amsfonts,amsthm,amssymb}
\usepackage{mathtools}
\usepackage{bbold}
\usepackage{palatino}
\usepackage{eurosym}
\usepackage{enumerate}
\usepackage{ulem}
\usepackage[toc,page]{appendix}
\usepackage[inline]{enumitem}
\usepackage[%
colorlinks=true,
pdfborder={0 0 0},
linkcolor=red
]{hyperref}
\usepackage[top=2.5cm, left=2.5cm, right=2.5cm, bottom=2.5cm]{geometry}
\usepackage[square,numbers]{natbib}
\usepackage[etex=true,export]{adjustbox}
\numberwithin{equation}{section}
\usepackage{xr}
\usepackage{import}

\title{An individual-based model to explore the impact of psychological stress on immune infiltration into tumour spheroids}
\author[1,2]{Emma Leschiera\thanks{Corresponding author. Email: emma.leschiera@devinci.fr}}

\author[3]{Gheed Al-Hity}
\author[3]{Melanie S. Flint}

\author[4]{Chandrasekhar Venkataraman}

\author[5]{Tommaso Lorenzi} 
\author[6]{Luis Almeida}
\author[6,7]{Chloe Audebert}
\affil[1]{\small Léonard de Vinci Pôle Universitaire, Research Center, 92 916 Paris La Défense, France}
\affil[2]{\small Univ. Bordeaux, CNRS, INRIA, Bordeaux INP, IMB, UMR 5251, F-33400 Talence, France}
\affil[3]{School of Applied Sciences, University of Brighton, Centre for Stress and Age-related Diseases, Moulsecoomb, Brighton, BN2, 4GJ, UK.}
\affil[4]{School of Mathematical and Physical Sciences, University of Sussex, Department of Mathematics, Falmer, Brighton, BN1 9QH, UK.}

\affil[5]{Department of Mathematical Sciences ``G. L. Lagrange'', Politecnico di Torino, 10129 Torino, Italy.}
\affil[6]{Sorbonne Universit\'e, CNRS, Universit\'e de Paris, Laboratoire Jacques-Louis Lions UMR 7598, 75005 Paris, France.}
\affil[7]{Sorbonne Universit\'e, CNRS, Institut de biologie Paris-Seine (IBPS), Laboratoire de Biologie Computationnelle et Quantitative UMR 7238, 75005 Paris, France.}

\date{\vspace{-5ex}}

\begin{document}
		\maketitle
		\begin{abstract}
			In recent \textit{in vitro} experiments on co-culture between breast tumour spheroids and
			activated immune cells, it was observed that the introduction of the stress hormone cortisol resulted in a decreased immune
			cell infiltration into the spheroids. Moreover, the presence of cortisol deregulated the normal levels of the pro- and anti-inflammatory cytokines
			IFN-$\gamma$ and IL-10.
			We present an individual-based model to explore the interaction dynamics between tumour and immune cells under psychological  stress conditions. With our model, we explore the processes underlying the emergence of different levels of immune infiltration, with particular focus on the biological mechanisms regulated by IFN-$\gamma$ and IL-10. 
			The set-up of numerical simulations is defined to mimic the scenarios considered in the experimental study.  Similarly to the experimental quantitative analysis, we compute a score that quantifies the level of immune cell infiltration into the tumour. 
			The results of numerical simulations indicate that the motility of immune cells, their capability to infiltrate through tumour cells, their growth rate and the interplay between these cell parameters can affect the level of immune cell infiltration in different ways. 
			Ultimately, numerical simulations of this model
			support a deeper understanding  of the impact of biological stress-induced mechanisms on immune infiltration.
		\end{abstract}


		
		\section{Introduction} \label{Introduction}
		The ability of psychological stress to induce immune suppression is widely recognised \citep{coe2007psychosocial,morey2015current,seiler2020impact}, but the mechanisms underlying the effects of psychological stress on the adaptive immune response during tumour progression are not completely understood. 
		
		There has been increasing interest in detailing the mechanistic role that psychological stress may play in the context of initiation and progression of cancer. In particular, it has been reported that psychological stress positively influences carcinogenesis through mechanisms that promote proliferation, angiogenesis and metastasis, as well as mechanisms that protect tumour cells from apoptosis ~\citep{lee2009surgical,nilsson2007stress}. The negative role played by psychological stress on the immune system has also been documented. 
		Using a pre-clinical mouse model, in~\citep{budiu2017restraint} the authors have shown that psychological stress has a negative impact on T cell numbers and activation, as evidenced by a decrease in the numbers of CD8+ and CD3+CD69+ T cells.  
		
		In~\citep{al2021integrated}, the authors developed a 3D \textit{in vitro} model to explore the effects of
		the stress hormone cortisol on
		immune cell infiltration into tumour spheroids. 
		Using two independent image-based algorithms, they quantified the effects of cortisol on immune infiltration, which was assessed by counting the number of immune cells within the
		tumour spheroid boundary. 
		The results from
		this model recapitulated the conclusions of~\citep{budiu2017restraint}, by showing that cortisol triggered a reduction in immune infiltration levels. 
		
		The mixture of cytokines produced in the tumour-microenvironment plays a key role in tumour progression~\citep{dranoff2004cytokines}. Pro-inflammatory cytokines that are released in response to infection can inhibit tumour development and progression. Alternatively, tumour cells can produce anti-inflammatory cytokines that promote growth, attenuate apoptosis and facilitate metastasis. 
		In the experiments reported in~\citep{al2021integrated}, cortisol downregulated  IFN-$\gamma$ and upregulated IL-10. 
		IFN-$\gamma$ is a pro-inflammatory cytokine that stimulates immune response, through T cell trafficking in the tumour-microenvironment and infiltration~\citep{castro2018interferon,schiltz2002effects}, whereas IL-10 is an anti-inflammatory cytokine that inhibits immune response by reducing T cell proliferation~\citep{alhakeem2018chronic,couper200810}. 
		
		From a biological and medical perspective, it is difficult to investigate the connection between psychological stress, immune infiltration and the underlying molecular and cellular processes. The challenge lies in
		integrating theoretical and empirical knowledge to achieve a deeper understanding of the mechanisms
		and factors that contribute to inhibition of the anti-tumour immune response. In this context, mathematical models provide easy and cheap tools towards identifying dependencies between different biological phenomena and how
		these may affect the efficacy of the immune response on much shorter timescales  than laborious and expensive experiments. 
		Different aspects of the interaction dynamics between immune and tumour cells have been studied using different deterministic continuum models formulated as ordinary differential
		equations (ODEs)~\citep{de2003mathematical,kirschner1998modeling,kuznetsov1994nonlinear,luksza2017neoantigen}, integro-differential equations (IDEs)~\citep{delitala2013recognition,kolev2013numerical,lorenzi2015mathematical} and partial differential equations (PDEs)~\citep{atsou2020size, atsou2021size,matzavinos2004travelling,matzavinos2004mathematical}. These models usually describe the evolution of tumour and immune cell densities that depend on one or more independent variables, usually time and/or space. Such models are defined on the
		basis of cell population-level phenomenological assumptions, which may limit the amount of biological detail incorporated into the model. By using computational models, such as cellular-automaton (CA) models~\citep{bouchnita2017hybrid,ghaffarizadeh2018physicell}, hybrid PDE-CA models~\citep{almeida2022hybrid,leschiera2021mathematical} and individual-based models (IBMs)~\citep{christophe2015biased,kather2017silico,macfarlane2018modelling, macfarlane2019stochastic}, a mathematical representation
		of biological phenomena that are challenging to include in purely continuum models can be achieved. In fact, these models can be posed on a spatial domain, where cells spatially interact with each other according to a defined set of rules, which can collectively generate global emergent behaviours of tumour-immune cell competition. 
		
		In~\citep{leschiera2021mathematical}, we proposed an IBM that describes the earliest stages of tumour-immune competition. In this model, we included cytotoxic T lymphocytes (CTLs) and tumour cells, which interact in a two-dimensional domain under a set of rules describing cell division, migration via chemotaxis, cytotoxic killing of tumour cells by CTLs and immune evasion. 
		However, the model in~\citep{leschiera2021mathematical} does not consider the role played by psychological stress
		in immune infiltration and the influence of pro- and anti-inflammatory cytokines on tumour progression.   These aspects are addressed in the present work.
		
		In light of these considerations, and motivated by \textit{in vitro} experimental observations in co-culture between cancer
		spheroids and immune cells~\citep{al2021integrated}, in this paper we develop an IBM to study the effect of psychological stress on immune infiltration. 
		The model builds on our previous work~\citep{leschiera2021mathematical} and is calibrated to qualitatively reproduce, \textit{in silico}, the experimental results presented in~\citep{al2021integrated}.  As mentioned earlier, in this study the authors found that the introduction of cortisol in the co-culture resulted in a decrease in immune cell infiltration into tumour spheroids, as well as in the alteration of IFN-$\gamma$ and IL-10 levels. In our model,
		we assume that cells are exposed to psychological stress, and that this deregulates IFN-$\gamma$ and IL-10 levels. We explore the processes underlying the emergence of different levels of immune infiltration, with particular focus on biological mechanisms regulated by IFN-$\gamma$ and IL-10.  
		Based on one of the two image-based algorithms developed in~\citep{al2021integrated} to quantify immune infiltration, in our numerical simulations we compute a score to quantify the effects of psychological stress on immune infiltration.

		\section{Methods}
		\subsection{Summary of the experimental protocol employed in~\citep{al2021integrated}}
\begin{figure}[!t]
			\centering
			\includegraphics[width=15cm]{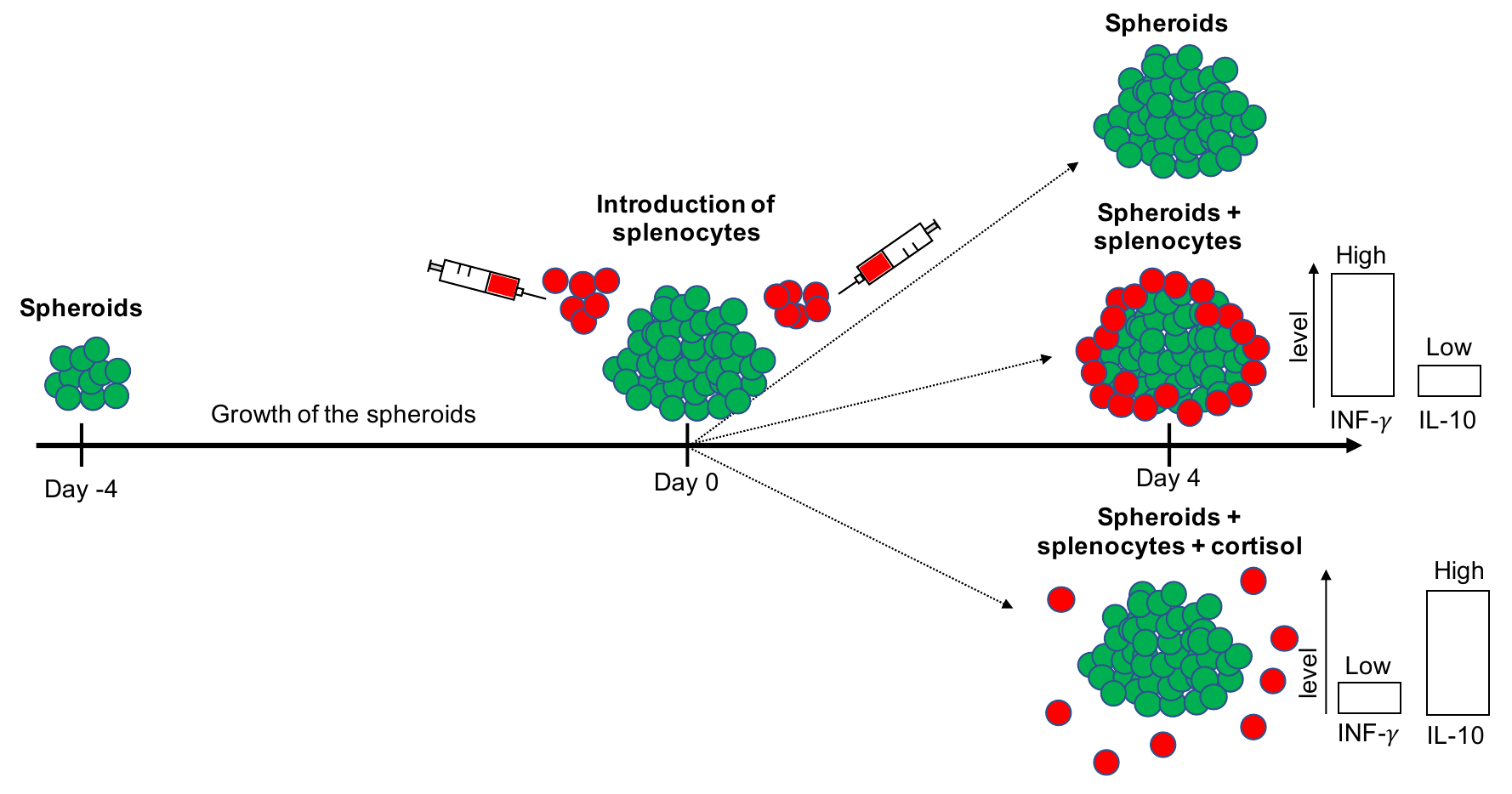}
			\caption{\textbf{Schematic summarising the experimental protocol employed in~\citep{al2021integrated}.} 
			}
			\label{ch4:experimentalStrategies}
		\end{figure}

		A summary of the experimental protocol employed in~\citep{al2021integrated} is provided below (further details can be found in~\citep{al2021integrated}). The mathematical model and numerical simulations will then be implemented accordingly, in order to facilitate comparison with the experiments. A schematic summarising the experimental procedure employed in~\citep{al2021integrated} is displayed in Fig.\ref{ch4:experimentalStrategies}.
		
		
		\paragraph{Growth of the spheroids} Spheroids from a murine triple
		negative 66CL4 breast cancer cell line were generated. It took 4 days for the spheroids to fully form, during which their area increased over time. 
		The cell line
		was seeded at different densities. 
		The seeding density with the
		largest area, and with the least variation over
		4 days of growth post culturing cells into spheroids, was chosen (\textit{cf.} Group-\textbf{1} in Fig.\ref{ch4:experimentalStrategies}). This was done to ensure that the size of the spheroids remained stable  and that changes in the spheroids were due to the infiltration of immune cells. 
		
		\paragraph{Introduction of splenocytes} After full spheroid generation, immune cells (splenocytes), containing activated T lymphocytes, were co-cultured with spheroids. 
		Cortisol was added to the co-culture and spheroids were later split into different groups. The groups relevant to our study are: Group-\textbf{1} containing spheroids only, Group-\textbf{2} comprising spheroids and splenocytes, and Group-\textbf{3} containing spheroids, splenocytes and cortisol. 
		\paragraph{Trafficking index measuring infiltration levels of immune cells} To test whether cortisol caused a reduction in immune infiltration in the co-culture, each group was imaged daily for 4 days.  From two imaged-based algorithms, two trafficking indices   were computed every day to measure immune cell infiltration into the spheroids \citep{al2021integrated} (\textit{i.e.} to quantify the number of immune cells within the  boundary of the tumour spheroids). Below we report on the implementation of the trafficking index that inspired the development of the score proposed in the present study to measure infiltration levels. Further details can be found in \citep{al2021integrated}.
		
		The \textit{classification-based trafficking index} (TIC) is an algorithm 
		in which first a machine learning algorithm is used to classify each pixel in the image into different groups obtaining three classes: background, tumour cell or immune cell. 
		The TIC is then based on the number of pixels classified as immune cells that are completely surrounded by pixels classified as tumour cells, divided by the total count of pixels surrounded by tumour cells. The resulting statistic yields a number in the interval $[0,1]$, with a larger TIC indicating a greater level of trafficking.  
		
		\paragraph{Investigating the effects of cortisol on immune infiltration} Over 4 days, the co-cultures in Group-\textbf{2} and Group-\textbf{3} were
		imaged and the TIC was computed from the corresponding images.
		
		It was found that, compared to the group with untreated spheroids and splenocytes only (\textit{i.e.} Group-\textbf{2}), the introduction of cortisol significantly reduced immune cell infiltration into the spheroids in Group-\textbf{3}. 
		Moreover, cortisol significantly reduced IFN-$\gamma$ levels and increased IL-10 levels.
		\label{Description of the experiments}
		
		\subsection{Modelling framework}
		Building upon our previous work~\citep{leschiera2021mathematical}, to reproduce the \textit{in vitro} results 
		presented in~\citep{al2021integrated}, we consider two cell types: tumour cells and immune cells. Although activated immune cells have been considered in the experiments, here we focus on CTLs only. We use a Cellular Potts (CP) approach and the CompuCell3D open-source software~\citep{izaguirre2004compucell} to model the interactions between these two cell types.
		
		Adhesive interactions between cells may affect the physical capability of CTLs to infiltrate through tumour cells. In this context, the choice of a CP model is of particular interest, since adhesive interactions between neighboring cells are represented through specific parameters, which describe the net adhesion/repulsion between cell membranes~\citep{izaguirre2004compucell} (see Supp.Mat.\ref{Details of computational model} for a detailed description of the implementation of our CP model). 
		
		Moreover, the CompuCell3D software easily allows for the visualisation of the results of numerical simulations. To carry out numerical simulations of the model, we consider both a 2D square spatial domain and a 3D cubic spatial domain. 
		
		At each time step, the states of the cells
		are updated according to the rules described below.
		\paragraph{Growth of tumour cells}
		We denote by $N_T(t)$ the number of tumour cells in the system at time $t$ 
		and we label each cell by an index $n=1,\dots, N_T(t)$.
		
		In the experimental setup described in Sec.~\ref{Description of the experiments}, spheroids are cultured for a sufficient time until they attain a stable size. To reproduce such dynamics, we let tumour cells proliferate until a maximum tumour size is attained, which corresponds to the carrying capacity of the population. With the selected parameter values, this process takes approximately 7 days. 
		
		At the initial time of simulations, we assume a certain number of tumour cells to 
		be tightly packed in a circular configuration positioned at the centre of the domain. 
		At each time-step, tumour cells grow at a rate drawn from a uniform distribution. 
		Mitosis occurs when tumour cells grow to a critical size and then
		divide. We refer the reader to our previous paper~\citep{leschiera2021mathematical} for a detailed description of the modelling strategy employed to represent cell division. 
		
		Tumour cells can
		die due to intra-population competition (\textit{i.e.} competition between tumour cells for limited space and resources), at a rate proportional to the total number of tumour cells. If tumour cells exhaust their lifespan (which is drawn from a uniform distribution when cells are
		created) then they die. Dead tumour cells are removed from the domain. 
		\paragraph{Introduction of CTLs}
		We denote by $N_C(t)$ the number of CTLs in the system at time $t$, and we label each of them by an index $m= 1,...,N_{C}(t)$.

		When introduced, CTLs are randomly distributed at the border of the spatial domain. 
		Once
			in the domain, CTLs grow and divide through mitosis according to rules similar to those used for tumour cells. CTLs die due to intra-population competition (\textit{i.e.} competition between CTLs for limited space and resources), at a rate proportional to the total number of CTLs.  A CTL can also die due to natural death when it reaches the end of its intrinsic lifespan, which is drawn from a uniform distribution when the cell is created.
		
		We assume that tumour cells at the border of the tumour (the region where cytokines and immune cells are more abundant \citep{boissonnas2007vivo}) secrete a chemoattractant. We assume that this chemoattractant represents  a mixture of different chemokines (e.g. CXCL9/10/11 \citep{galon2019approaches,gorbachev2007cxc}) produced by tumour cells and other cells in the tumour micro-environment (e.g. monocytes, endothelial cells, fibroblasts \citep{tokunaga2018cxcl9}), and that it triggers
			the movement of CTLs towards tumour cells.  A detailed description of the chemoattractant dynamics is given in Sup.Mat.\ref{Details of computational model}. CTLs are assumed to move up the gradient of the chemoattractant towards tumour cells.
		
		According to the experiments, CTLs are activated against tumour cells. Therefore, we suppose that, upon contact, CTLs can induce tumour cell death with a certain probability. We refer to this probability as the “immune success rate”. If tumour cells satisfy the conditions to be eliminated then they
		die. 
		\paragraph{Infiltration score} Similarly to the TIC proposed in~\citep{al2021integrated} to measure immune infiltration, in our work we define an `infiltration score'. This score allows us to quantify the level of CTL infiltration into the tumour. 
		Provided that there are tumour cells in the domain, we define the infiltration score as the number of CTLs surrounded by tumour cells, divided by the number of tumour cells and CTLs surrounded by tumour cells; that is, 
		\begin{equation}
			I(t) := \frac{\sum_{m=1}^{{N_C(t)}}\delta_{{m\in N_{CS}(t)}}}{\sum_{m=1}^{{N_C(t)}}\delta_{{m\in N_{CS}(t)}}+\sum_{n=1}^{N_T(t)}\delta_{{n\in N_{TS}(t)}}}.
			\label{infiltration score}
		\end{equation}
		In \eqref{infiltration score}, $\delta_{{m\in N_{CS}(t)}}=1$ if $m\in N_{CS}(t)$, and $\delta_{{m\in N_{CS}(t)}}=0$ otherwise, where $N_{CS}(t)$ denotes the set of indices of CTLs surrounded by tumour cells at time $t$.  Function $\delta_{{n\in N_{TS}(t)}}$ is defined in a similar way, where $N_{TS}(t)$ denote the set of indices of tumour cells surrounded by tumour cells at time $t$. In Compucell3D, these terms are handled by using specific functions which track neighbors of every cell (further details can
		be found in Sup.Mat.\ref{Details of computational model}).
		Note that $0\leq I(t) \leq 1$.
		
		\paragraph{Investigating the effects of psychological  stress on immune infiltration}
		Over 4 days, CTLs move via
		chemotaxis towards the tumour and infiltrate into it.
		
		In the experiments reported in  \citep{al2021integrated}, the spatio-temporal dynamics of cortisol,  IFN-$\gamma$ and IL-10   are not measured. The measurements  focus on the level of immune infiltration into tumour spheroids, as well as IFN-$\gamma$ and IL-10 amounts after 4 days of co-culture. 
			Therefore, to reduce the number of parameters and the complexity of the model, we do not explicitely model the dynamics of cortisol, IFN-$\gamma$ or IL-10. Through numerical simulations, we investigate  the effects of three parameters associated to IFN-$\gamma$ and IL-10, which we expect to play a key role in determining the infiltration of CTLs into the tumour. These three parameters are: the secretion rate of the chemoattractant by tumour cells, the
		"tumour cell-CTL adhesion strength" and the growth
		rate of CTLs. Below we detail how these three parameters are associated to IFN-$\gamma$ and IL-10.

		It has been shown that IFN-$\gamma$ induces the stimulation of various chemokines (\textit{e.g.} CXCL9/10/11) which drive the chemotactic movement of CTLs towards the tumour \citep{castro2018interferon,galon2019approaches, gorbachev2007cxc}. 
		Therefore, in our study, the role of IFN-$\gamma$ is investigated by varying the secretion rate of the chemoattractant by tumour cells.

		Moreover, IFN-$\gamma$ induces the expression of cellular adhesion molecules (\textit{e.g.} E-cadherin or ICAM-1), which enhance the infiltration of CTLs into the tumour \citep{harjunpaa2019cell, jorgovanovic2020roles}. Therefore, the role of IFN-$\gamma$ is also investigated by varying the "tumour cell-CTL adhesion strength" (TC-CTL adhesion strength). 
		This parameter refers to the CP parameter associated to the adhesion between tumour cell and CTL membranes. In our model, the TC-CTL adhesion strength regulates CTL ability to infiltrate through tumour cells. In particular, high values of this parameter facilitate the infiltration of CTLs through tumour cells, while low values lead the CTLs to accumulate at the margin of the tumour, without infiltrating into it. More details on the calibration of this parameter are provided in Sup.Mat.\ref{appendix5:graph}.

		Finally, IL-10 is an immunoregulatory cytokine that can attenuate inflammatory responses by suppressing CTL production and proliferation~\citep{alhakeem2018chronic,couper200810}. 
		Therefore, the effect of IL-10 is investigated by varying the growth rate of CTLs.
		
		In this work, we explore different scenarios. We suppose that in non-stressed conditions IFN-$\gamma$ levels are high, IL-10 levels are low, and CTLs infiltrate into the tumour. In stressed conditions instead, we suppose that IFN-$\gamma$ levels decrease and  IL-10 levels increase, leading to a decreased CTL infiltration. 
		By considering a range of values of these three parameters, we explore their impact on tumour-immune dynamics independently and together, assessing their influence on immune infiltration in a controlled manner.
		\label{The mathematical model}
		\section{Preliminary results in 2D and 3D}
		\label{ch4:Numerical simulations and preliminary results}
		In this section, the results of preliminary numerical simulations of the model in 2D and 3D are presented, which will be used to guide the simulations leading to the main results presented in Sec.~\ref{ch4:Main results}. 
		
		Given the stochastic nature of the model, all the results we present in this section and in Sec.~\ref{ch4:Main results} are obtained by averaging over $5$ simulations, which were 
		carried out using the parameter values reported in Tab.\ref{ch4:table1} and Tab.\ref{ch4:table2}. It should be noted that the standard deviation between these 5 simulations is relatively small which leads us not to increase the number of simulations (and their relative computational cost). A lower number of replicates would  not allow to check the robustness of the results. Full details of model implementation and model parameterisation are provided in Sup.Mat.\ref{Details of computational model} and Sup.Mat.\ref{appendix5:graph}, respectively. Files to run a simulation example of the model with Compucell3D \cite{izaguirre2004compucell} are available at \url{https://plmlab.math.cnrs.fr/leschiera/roleofstress}.

		\subsection{Tumour development in the absence of CTLs}
		\label{ch4:preliminary scenario}
\begin{figure}[!t]
			\centering
			\includegraphics[width=15.5cm]{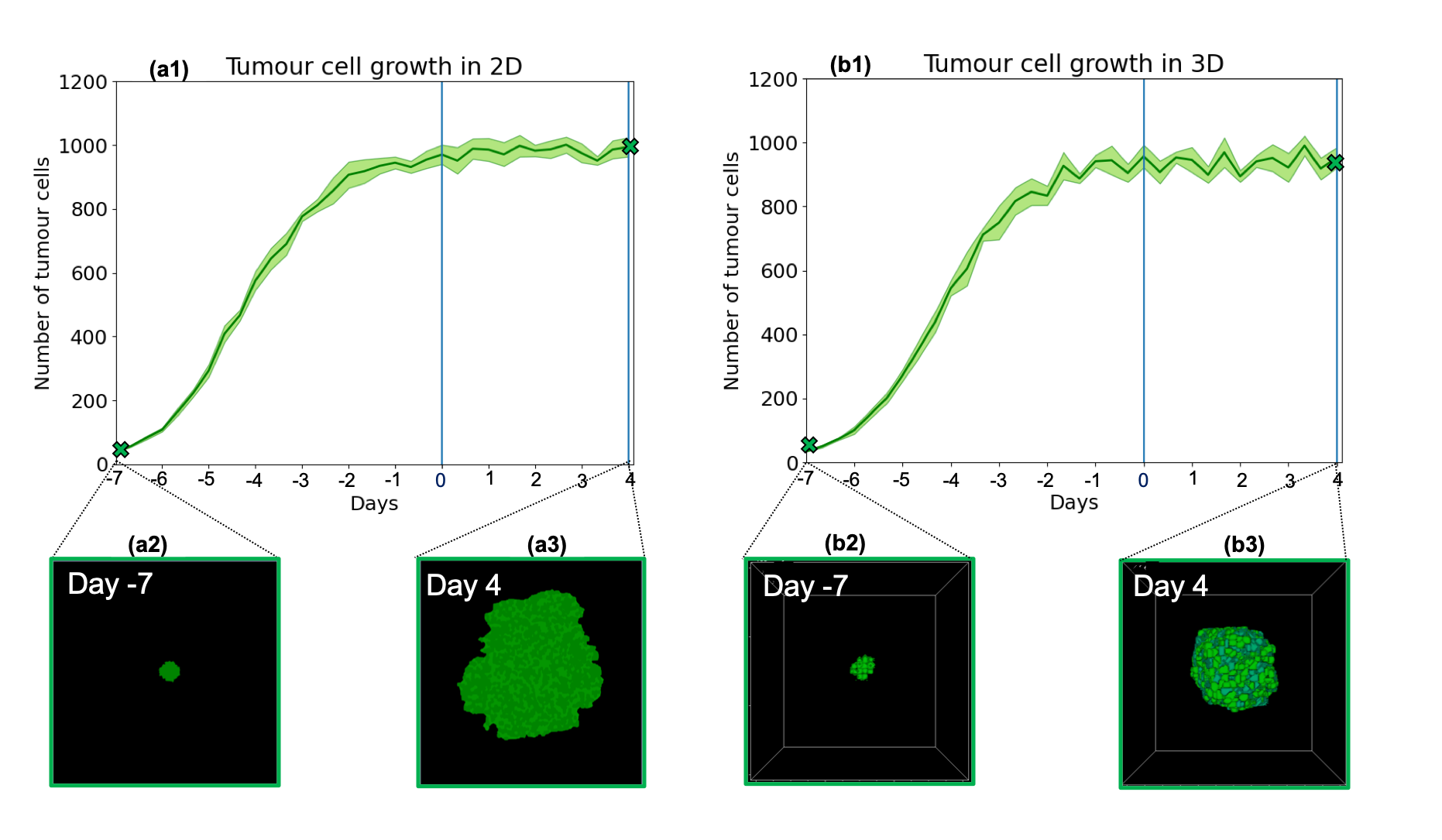}
			\caption{\textbf{Tumour development in the absence of CTLs.} Panels \textbf{(a1)} and \textbf{(b1)} display the time evolution of the tumour cell number in the absence of CTLs in 2D and 3D, respectively. These results correspond to the average over 5 simulations and the shaded areas indicate $+/-$ standard deviation. Panels \textbf{(a2)-(a3)} and \textbf{(b2)-(b3)} display samples of the initial and final spatial distributions of tumour cells in 2D and 3D, respectively. The parameter values used to carry out numerical simulations are provided in Tab.\ref{ch4:table1} and Tab.\ref{ch4:table2}.
			}
			\label{ch4:tumour_no_T_cells}
		\end{figure}

		We first establish a preliminary scenario where tumour cells grow, divide and die according to the rules described in Sec.~\ref{The mathematical model}, in the absence of CTLs.  At the initial time of simulations, 35 tumour cells are placed in the domain. 
		More details about the definition of the model initial conditions are given in Sup.Mat.\ref{appendix5:graph}. We carry out numerical simulations for $11$ days (which we count from day -7 to day 4).
		The plots in Fig.\ref{ch4:tumour_no_T_cells}\textbf{(a1)-(b1)} show the time evolution of the tumour cell number in 2D and 3D, while Fig.\ref{ch4:tumour_no_T_cells}\textbf{(a2)-(a3)} and Fig.\ref{ch4:tumour_no_T_cells}\textbf{(b2)-(b3)} display samples of initial and final spatial distributions of tumour cells 2D and 3D, respectively. The tumour growth is of logistic type, as expected due to the rules that govern tumour cell division and death. In more detail, as shown by Fig.\ref{ch4:tumour_no_T_cells}\textbf{(a1)-(b1)}, the number of tumour cells increases from day -7 to day 0. Around day 0, it reaches the carrying capacity. 
		The number of tumour cells at carrying capacity is similar to the seeding density chosen in ~\citep{al2021integrated} (\textit{cf.} Fig.1 for 66CL4 spheroids in~\citep{al2021integrated}). From day 0 to day 4, the tumour cell number fluctuates around the carrying capacity. 
		
		These simulations allowed us to calibrate the model parameters related to tumour cells in order to qualitatively reproduce the growth of the spheroids obtained in the experiments. The other simulations were carried out keeping the values of these parameters fixed and equal to those used for these simulations.
		
		\subsection{Control scenario: CTL infiltration in non-stressed conditions}
		\label{Control scenario: CTL infiltration in non-stressed conditions}
		In the experimental results presented by~\citep{al2021integrated}, in the absence of cortisol, immune cells are able to infiltrate the tumour spheroids. Here we verify the ability of our model to reproduce such dynamics by exploring the infiltration of CTLs into the tumour over $4$ days. For these simulations, the initial number of tumour cells is set at carrying capacity, whereas 
		150 CTLs are introduced in the domain. The values of the parameters related to CTLs are chosen so as to qualitatively reproduce the interaction dynamics between spheroids and immune cells in non-stressed conditions presented in~\citep{al2021integrated}.
		The parameters related to IFN-$\gamma$ and IL-10 levels are set to baseline values (\textit{i.e.} non-stressed conditions). In particular, we let tumour cells secrete the chemoattractant at a high rate, CTLs grow at their normal rate and display a high capability to infiltrate the tumour cell population  (\textit{i.e.} we consider a sufficiently high value for the TC-CTL adhesion strength). In order to gain a deeper understanding of the effects produced by the three aforementioned parameters on immune infiltration, 
		for the moment we simplify our model by assuming that CTLs are not able to eliminate tumour cells (\textit{i.e.} the immune success rate is set equal to 0). The full model with an immune success rate greater than 0 will be considered in Sec.~\ref{Increasing the immune success rate has an impact on the infiltration score only when the TC-CTL adhesion strength is high}.

\begin{figure}[!t]
			\centering
			\includegraphics[width=16cm]{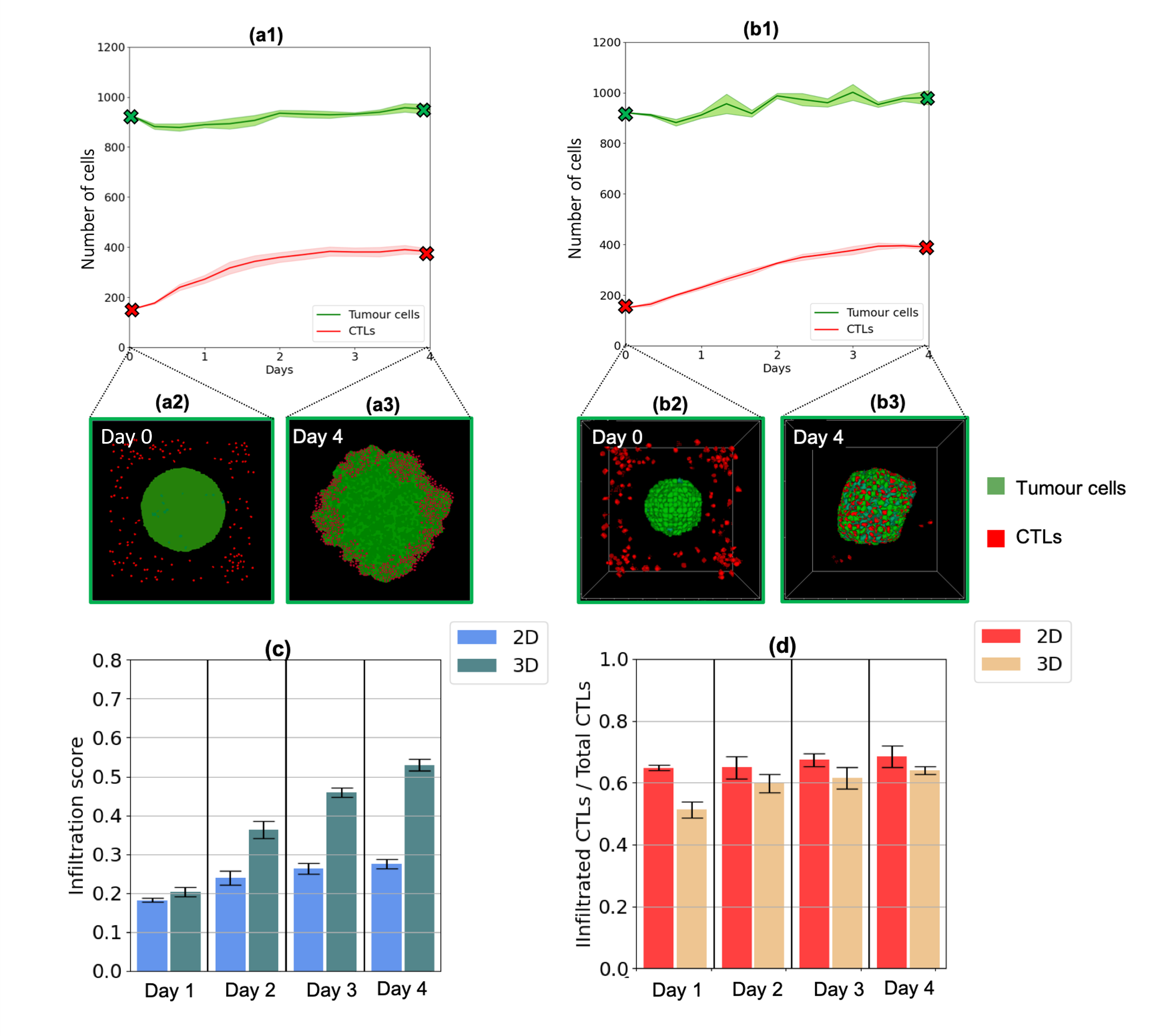}
			\caption{\textbf{Control scenario: CTL infiltration in non-stressed conditions.} Panel \textbf{(a1)-(b1)} display, respectively,  the time evolution of the numbers of tumour cells (in green) and CTLs (in red) in 2D and 3D  for a choice of parameter values that results in the infiltration of CTLs into the tumour. These results correspond to the average over 5 simulations and the shaded area indicates $+/-$ standard deviation. Panels \textbf{(a2)-(a3)} and \textbf{(b2)-(b3)} display samples of the initial and final spatial distribution of tumour cells (in green) and CTLs (in red) in 2D and 3D, respectively. Panel \textbf{(c)} displays the corresponding
				average value of the infiltration score, computed via \eqref{infiltration score}, at different times of the 2D and 3D simulations. The error lines
				represent the standard deviation between 5 simulations. Panel \textbf{(d)} displays the ratio between the corresponding average number of infiltrated CTLs and the total number of CTLs at the
				end of 2D and 3D simulations. The error lines represent the standard deviation between 5 simulations. The parameter values used to carry out numerical simulations are provided in Tab.\ref{ch4:table1} and Tab.\ref{ch4:table2}.
			}
			\label{ch4:baselineTOT}
		\end{figure}

		The plots in Figs.\ref{ch4:baselineTOT}\textbf{(a1)}-\textbf{(b1)} show the time evolution of the number of tumour cells  and CTLs in 2D and 3D, while Figs.\ref{ch4:baselineTOT}\textbf{(a2)}-\textbf{(a3)} and Figs.\ref{ch4:baselineTOT}\textbf{(b2)}-\textbf{(b3)} display samples of initial and final spatial distributions of tumour cells and CTLs in 2D and 3D, respectively.
		The choice of parameter values corresponding to these figures results in the infiltration of CTLs into the tumour.
		The plots in Figs.\ref{ch4:baselineTOT}\textbf{(c)}-\textbf{(d)} display, respectively, the corresponding average value of the infiltration score, computed via~\eqref{infiltration score}, and the average number of infiltrated CTLs over 4 days. Both in 2D and 3D, as soon as CTLs are introduced in the domain, they move towards the tumour and infiltrate it. Fig.\ref{ch4:baselineTOT}\textbf{(c)} indicates that the infiltration score increases over time, both in 2D and 3D. In 2D its value tends to saturate between day 3 and day 4. Moreover, the value of the infiltration score in the 3D setting is larger than in the 2D case. 
		Note that, in 2D, the mean value of the infiltration score obtained at day 4 of simulations is similar to the mean value of the TIC computed in~\citep{al2021integrated}, when cortisol was not introduced in the co-culture (\textit{cf.} Fig.5 in~\citep{al2021integrated}). 
		Fig.\ref{ch4:baselineTOT}\textbf{(d)} demonstrates that, in 2D, most of the CTLs infiltrate the tumour already at day 1, as the average number of infiltrated CTLs increases only slightly between day 1 and 4. On the other hand, in 3D, CTLs seem to be slightly slower in moving towards the tumour. However, the average number of infiltrated CTLs at the end of numerical simulations is similar in the two settings.  Finally, as shown by Figs.\ref{ch4:baselineTOT}\textbf{(a1)-(b1)}, and as expected on the basis of the rules that govern tumour cell and CTL growth and death, both in 2D and 3D, over time the number of tumour cells fluctuates around the carrying capacity, while CTL number increases until it reaches a saturation value. This result indicates that the changes in the tumour surface and volume observed in Fig.\ref{ch4:baselineTOT}\textbf{(a3)} and Fig.\ref{ch4:baselineTOT}\textbf{(b3)} are due to the infiltration of CTLs into the tumour.


		\section{Main results}
		\label{ch4:Main results}
		In this section we explore the effects of psychological stress on immune infiltration. 
		To do so, first we decrease the secretion rate of the chemoattractant and the TC-CTL adhesion strength. These two parameters are associated with decreased levels of IFN-$\gamma$. Next, for each scenario considered, we decrease the growth rate of CTLs, which is associated with increased levels of IL-10. The initial number and position of tumour cells and CTLs is kept equal to that used in the control scenario.
		
		Conducting baseline numerical simulations in 3D provided valuable insights into potential disparities between calculating the infiltration score on 2D and 3D images. Nevertheless, since in the experiments reported in \citep{al2021integrated} the infiltration score is computed on 2D images, we now carry out 2D simulations only, also because they require much less computational time than the corresponding 3D simulations.

		In this section, we report on results obtained by varying the values of the three aforementioned parameters 
		while the other parameters are kept equal to the values used in the previous section. For each scenario, the infiltration score is computed via \eqref{infiltration score}. 
		
		\subsection{Decreasing the secretion rate of the chemoattractant and the TC-CTL adhesion strength reduces the infiltration of CTLs into the tumour}
		\label{Decreasing the secretion rate of the chemoattractant reduces the infiltration of CTLs into the tumour}
		To investigate how immune infiltration is affected by IFN-$\gamma$ levels in the domain, we start
		by comparing the control scenario of Sec. \ref{Control scenario: CTL infiltration in non-stressed conditions} with  scenarios  in which the values of the secretion rate of the chemoattractant and of the TC-CTL adhesion strength are reduced (\textit{cf.} Tab.\ref{ch4:table1} and Tab.\ref{ch4:table2}).

\begin{figure}[!t]
			\centering
			\includegraphics[width=15cm]{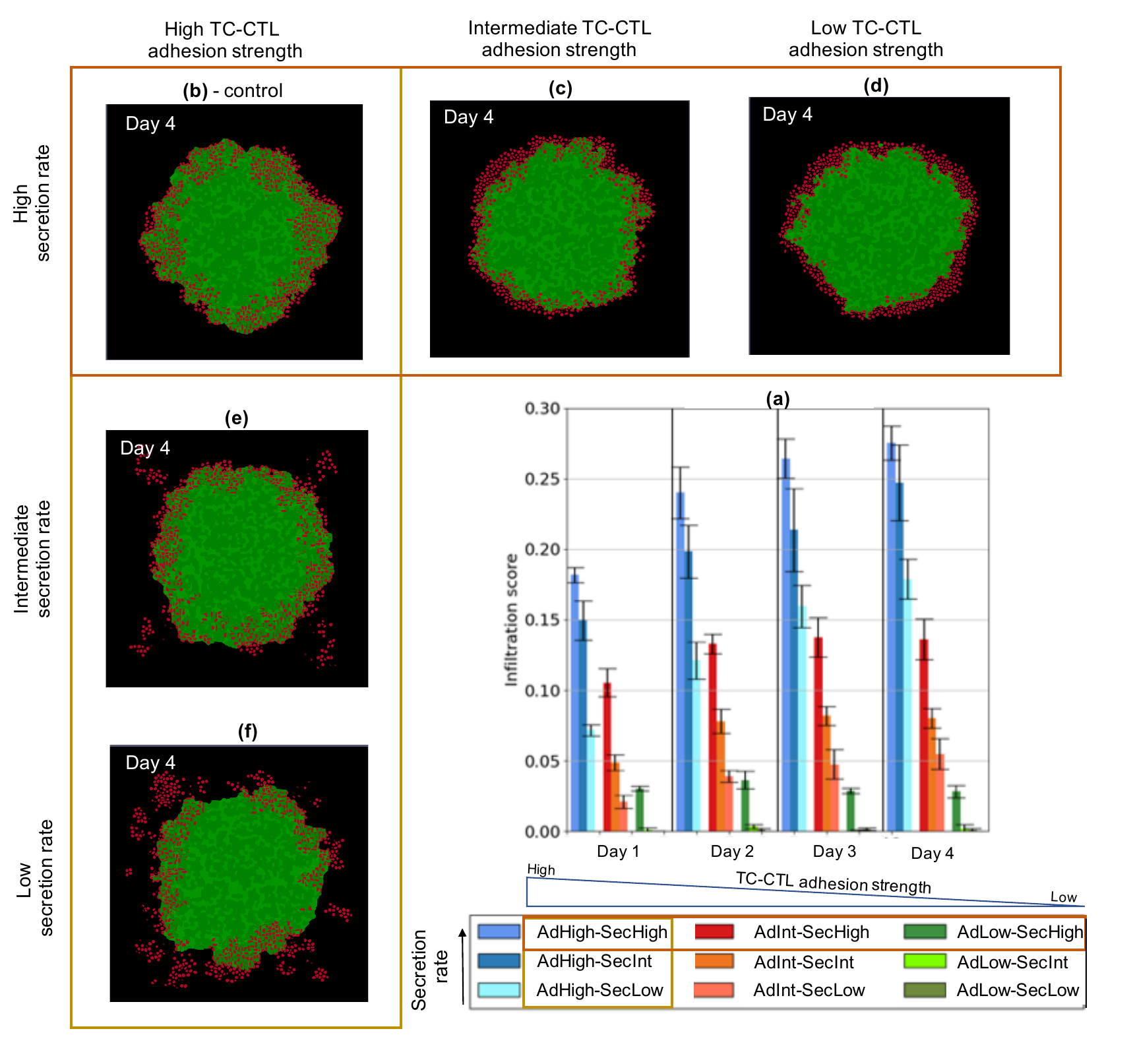}
			\caption{\textbf{Decreasing the secretion rate of the chemoattractant  and the TC-CTL adhesion strength reduces the infiltration of CTLs into the tumour.} Panel \textbf{(a)} displays the
				average value of the infiltration score, computed via \eqref{infiltration score}, for different values of the secretion rate of the chemoattractant and the TC-CTL adhesion strength, at different times of the simulations. The error lines
				represent the standard deviation between 5 simulations. Panel \textbf{(b)} displays a sample of the final spatial distribution of tumour cells (in green) and CTLs (in red) for the parameter values considered in the control scenario.  Panels \textbf{(c)}-\textbf{(d)} display similar plots for 2 different values of the TC-CTL adhesion strength. Panels \textbf{(e)}-\textbf{(f)} display similar plots for 2 different values of the secretion rate of the chemoattractant. The parameter values used to carry out numerical simulations are provided in Tab.\ref{ch4:table1} and Tab.\ref{ch4:table2}.
			}
			\label{ch4:mainres1}
		\end{figure}

		Fig.\ref{ch4:mainres1}\textbf{(a)} displays the average value of the infiltration score at different times of the simulations, for high, intermediate and low values of the secretion rate of the chemoattractant and the TC-CTL adhesion strength. 
		This figure shows that both parameters affect the infiltration of CTLs into the tumour, as the infiltration score decreases as soon as one of the two parameters is reduced. In addition, when the
		TC-CTL adhesion strength is sufficiently high, decreasing the secretion rate of the chemoattractant considerably reduces the infiltration score. On the other hand, for sufficiently  low values of the
		TC-CTL adhesion strength, decreasing the secretion rate of the chemoattractant does not have an impact on the infiltration score, as its value is already small. Taken together, these results suggest that the secretion rate of the chemoattractant has an impact on T cell infiltration only when CTLs display a sufficiently high capability to infiltrate through tumour cells. 
		
		Then, we analyse the spatial cell distributions observed at the end of simulations. Figs.\ref{ch4:mainres1}\textbf{(c)}-\textbf{(d)} show samples of the final spatial distributions of tumour cells and CTLs for intermediate and low values of the TC-CTL adhesion strength. Figs.\ref{ch4:mainres1}\textbf{(e)}-\textbf{(f)} show similar plots for intermediate and low values of the secretion rate of the chemoattractant. 
		These plots are to be compared with the one in Fig.\ref{ch4:mainres1}\textbf{(b)}, which displays the final spatial distributions of tumour cells and CTLs obtained in the control scenario. In particular, Figs.\ref{ch4:mainres1}\textbf{(b)}-\textbf{(d)} show that decreasing  the TC-CTL adhesion strength leads to scenarios in which CTLs accumulate around the tumour, because  the secretion rate of the chemoattractant is high, but they do not infiltrate into it. The calculation of the infiltration score defined via \eqref{infiltration score} only takes into account CTLs infiltrated into the tumour but not the ones surrounding it. Therefore, the infiltration score decreases. On the other hand, Figs.\ref{ch4:mainres1}\textbf{(b)}-\textbf{(e)}-\textbf{(f)} indicate that decreasing the secretion rate of the chemoattractant leads to scenarios in which CTLs away from the tumour are not sensitive to the gradient of the chemoattractant and, therefore, do not move towards the tumour. The more CTLs are not sensitive to the chemoattractant and do not infiltrate the tumour, the more the infiltration score decreases.
		
		Taken together, these results qualitatively reproduce key experimental findings presented in~\citep{al2021integrated}, which indicated that cortisol reduced IFN-$\gamma$ levels and led also immune infiltration to reduce. The modelling assumption underlying these computational results may provide the following theoretical explanation for such behaviour. Since IFN-$\gamma$ may affect both CTL movement and infiltration capabilities, deregulation of IFN-$\gamma$ levels inhibits CTL ability to migrate towards the tumour and to infiltrate through tumour cells. The interplay between these mechanisms results in a progressive reduction of immune infiltration levels.
		
		\subsection{Decreasing the growth rate of CTLs reduces the number of infiltrated CTLs}
		We further investigate the effects of psychological stress on immune infiltration by exploring the role played by IL-10. For these simulations, we consider the same parameter values used in the previous subsection but we reduce the value of the growth rate of CTLs (\textit{cf.} Tab.\ref{ch4:table1} and Tab.\ref{ch4:table2}).
		
\begin{figure}[!t]
			\centering
			\includegraphics[width=16.5cm]{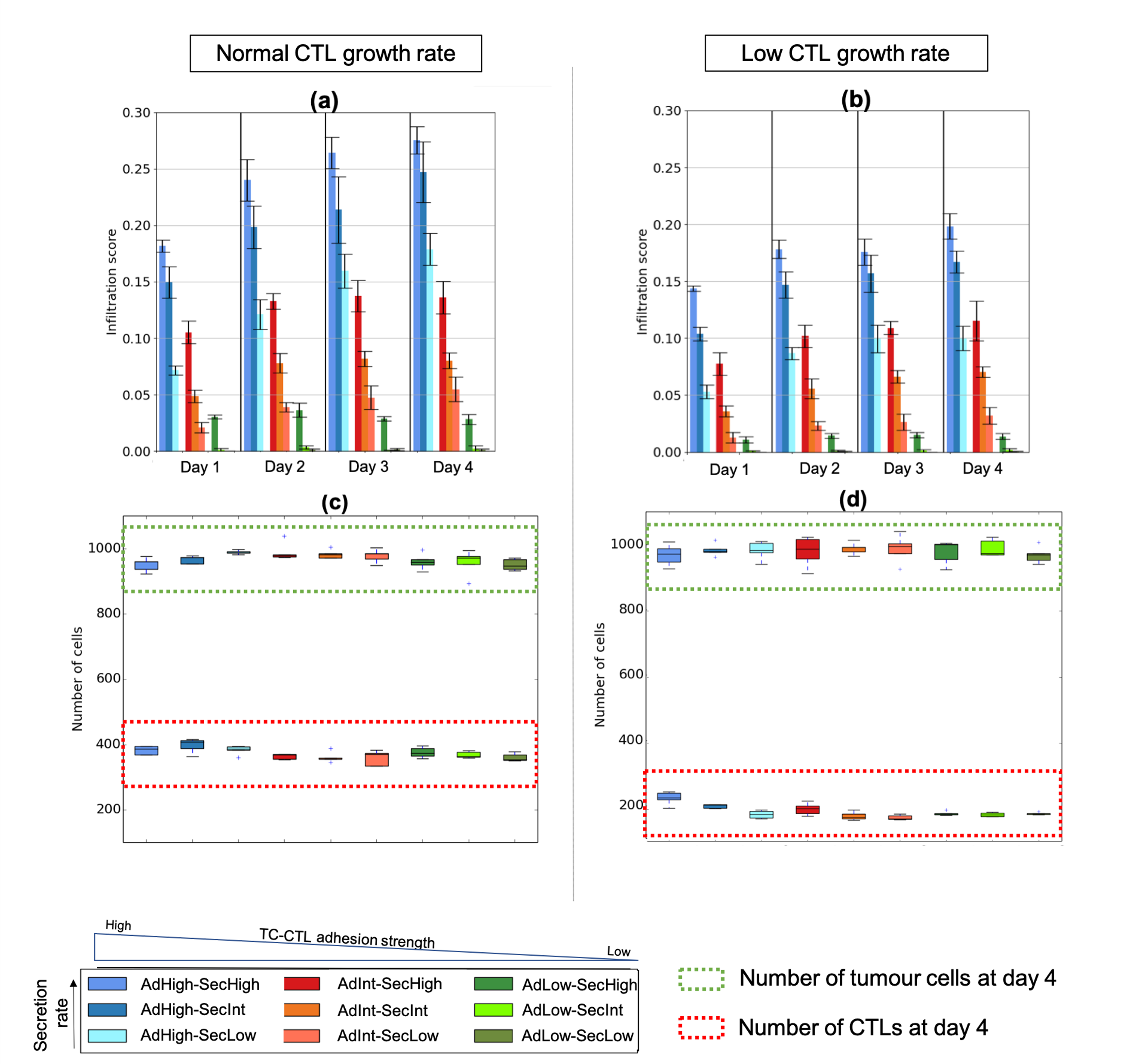}
			\caption{\textbf{Decreasing the growth rate of CTLs reduces the number of infiltrated CTLs.} Plots in panels \textbf{(a)}-\textbf{(b)} display the
				average value of the infiltration score, computed via \eqref{infiltration score}, for different values of the secretion rate of the chemoattractant and the TC-CTL adhesion strength, and at different times of the simulations. In panel \textbf{(a)} CTLs grow at their normal rate, while their growth rate is decreased in panel \textbf{(b)}. The error lines
				represent the standard deviation between 5 simulations. Panels \textbf{(c)}-\textbf{(d)} display the corresponding number of tumour cells and CTLs at the end of simulations (corresponding to day 4 of the experiments). The parameter values used to carry out numerical simulations are provided in Tab.\ref{ch4:table1} and Tab.\ref{ch4:table2}.
			}
			\label{ch4:mainres2}
		\end{figure}

		Figs.\ref{ch4:mainres2}\textbf{(a)}-\textbf{(b)} show a comparison between the infiltration score obtained when the effects of IL-10 are not considered (\textit{i.e.} for a normal value of the CTL growth rate), and the one obtained when the effects of IL-10 are considered (\textit{i.e.} when the CTL growth rate is reduced). Figs.\ref{ch4:mainres2}\textbf{(c)}-\textbf{(d)} also compare the number of tumour cells and CTLs at the end of numerical simulations (\textit{i.e.} at day 4 of the experiments) for the two scenarios considered. 
		Comparing the results of Fig.\ref{ch4:mainres2}\textbf{(a)} with those displayed in Fig.\ref{ch4:mainres2}\textbf{(b)}, we see that, similarly to the results observed in the previous subsection, decreasing the growth rate of CTLs reduces the infiltration score only when the TC-CTL adhesion strength is sufficiently high. 
		However, when  the TC-CTL adhesion strength is sufficiently low, decreasing the growth rate of CTLs does not have an impact on the infiltration score, as its value is already small.  As shown by Figs.\ref{ch4:mainres2}\textbf{(c)}-\textbf{(d)}, decreasing  the growth rate of CTLs leads to a decreased number of CTLs at the end of simulations, while the final number of tumour cells remains similar in the two scenarios.
		
		If we assume that high levels of IL-10 inhibit CTL growth, the outputs of our model indicate that, as expected, decreasing the proliferation rate of CTLs diminishes the number of CTLs in the domain. Moreover, if CTLs display a sufficiently high capability to infiltrate through tumour cells, we observe a reduction in the number of infiltrated CTLs (\textit{i.e.} the infiltration score decreases). On the other hand, if CTLs have a low capability to infiltrate through tumour cells, decreasing the proliferation rate of CTLs does not affect the infiltration score, as the number of infiltrated CTLs is already low. This suggests that high levels of IL-10 decrease immune infiltration only when CTLs display a sufficiently high capability to infiltrate through tumour cells, that is, when  IFN-$\gamma$ levels are sufficiently high.
		
		\subsection{Relationship between IFN-$\gamma$ and IL-10 levels and infiltration score}
		The results presented in the previous subsections summarise how scenarios corresponding to different levels of CTL infiltration into the tumour can emerge under sample combinations of the values of the secretion rate of the chemoattractant, the TC-CTL adhesion strength and the CTL growth rate.  We now undertake a more comprehensive investigation of the relationship between these parameters and the infiltration score.
			
			In order to do this, we perform numerical simulations holding all parameters constant but considering a broader range of values of the secretion rate of the chemoattractant by tumour cells, the tumour cell-CTL adhesion strength and the growth rate of CTLs.  For each pair of parameters,  the third parameter is set to its baseline value. For each scenario considered,  we determine the final value of the infiltration score computed via \eqref{infiltration score}. The results obtained are summarised in the heat maps of Fig.\ref{ch4:heatmap}. 
		
		As shown by the green regions on the bottom side of the two heat maps of Fig.\ref{ch4:heatmap}\textbf{(a)-(b)}, for sufficiently small values of the tumour cell-CTL adhesion strength, the immunoscore is relatively low (<0.1) independently of the value of the chemoattractant secretion rate and the CTL growth rate. This is due to the fact that, independently of their number or their sensitivity to the chemoattractant, CTLs accumulate around the tumour, but they might not infiltrate into it. On the other hand, when low values of the chemoattractant secretion rate  and the growth rate of CTLs are considered, but the baseline (\textit{i.e.} high) value of the tumour cell-CTL adhesion strength is considered (\textit{cf.} Fig.\ref{ch4:heatmap}\textbf{(c)}), the infiltration score is larger than in the two previous scenarios. 
		
		The orange-red regions of the heat maps of Fig.\ref{ch4:heatmap} indicate that there are several possible parameter ranges giving rise to an intermediate infiltration score (between 0.1 and 0.2). The first and second ones correspond to intermediate values of the tumour cell-CTL adhesion strength along with either intermediate to low  values of the chemoattractant secretion rate (\textit{cf.} Fig.\ref{ch4:heatmap}\textbf{(a)}) or normal to low  values of the CTL growth rate (\textit{cf.} Fig.\ref{ch4:heatmap}\textbf{(b)}). The third parameter range corresponds to low to normal values of the CTL growth rate along with intermediate to low values of the chemoattractant secretion rate (\textit{cf.} Fig.\ref{ch4:heatmap}\textbf{(c)}). 
		
		Finally, as shown by the light blue regions on the top-left side of Fig.\ref{ch4:heatmap}\textbf{(a)-(b)}, for high values of the three parameters, which correspond to the baseline parameters chosen in the control scenario, the value of the immunoscore is relatively high (>0.2). Fig.\ref{ch4:heatmap}{\textbf{(c)} shows that a relatively high immunoscore can be obtained also for lower values of the CTL growth rate (resp. chemoattractant secretion rate), provided that the chemoattractant secretion rate (resp. CTL growth rate) is large enough. Moreover, these results also show that increasing the growth rate of CTLs or the chemoattractant secretion rate to values higher than those considered in the control scenario does not always increase the infiltration score. 
			\begin{figure}[!t]
				\centering
				\includegraphics[width=14.5cm]{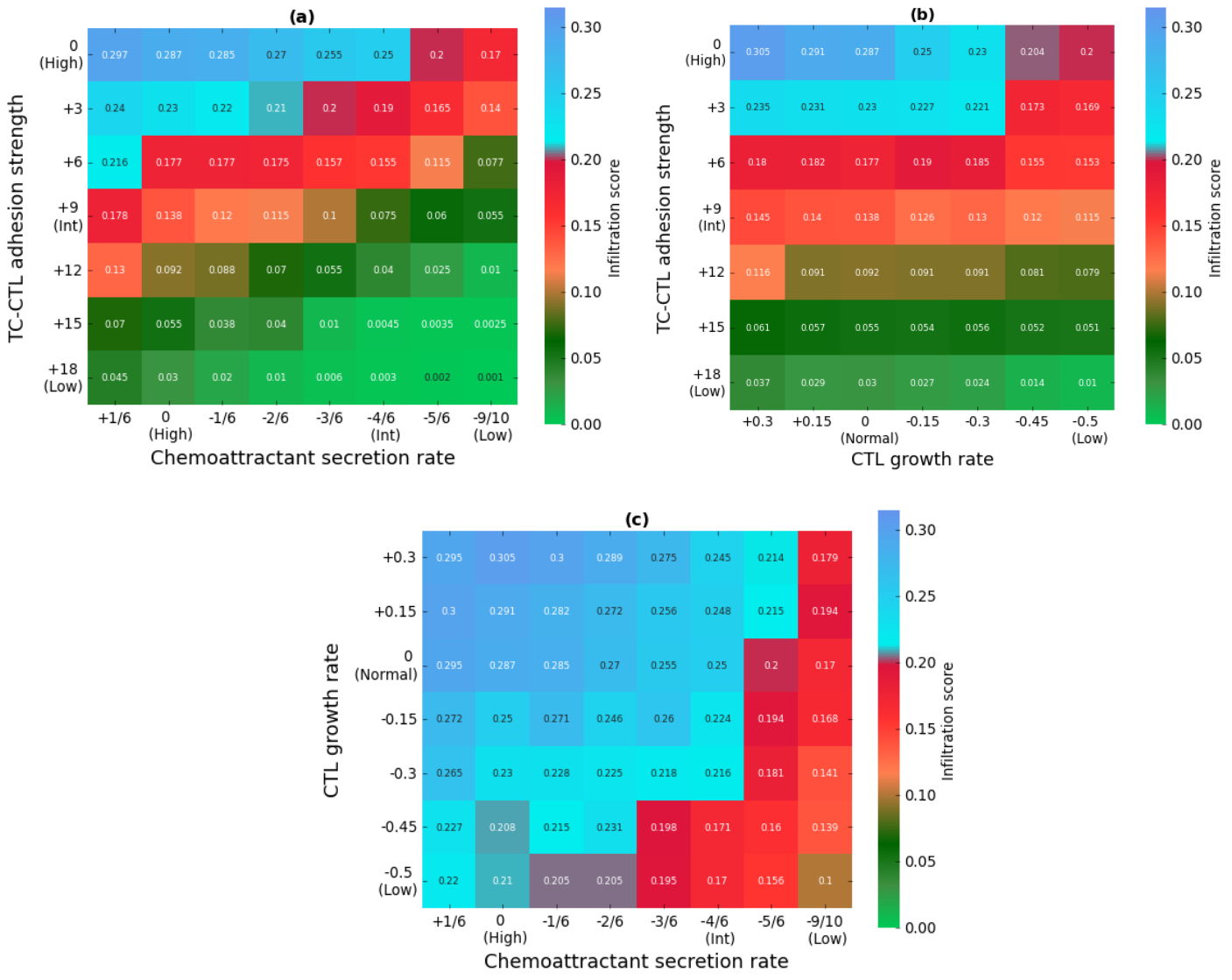}
				\caption{\textbf{Relationship between IFN-$\gamma$ and IL-10 levels and infiltration score.} Panel \textbf{(a)} displays the
					average value of the infiltration score, computed via \eqref{infiltration score}, for different values of the secretion rate of the chemoattractant and the TC-CTL adhesion strength at the end of the simulations. Panel \textbf{(b)} displays a similar plot for different values of the TC-CTL adhesion strength and the CTL growth rate. Panel \textbf{(c)} displays a similar plot for different values of the secretion rate of the chemoattractant and the CTL growth rate. Results are shown for variations of: +$\frac{1}{6}, -\frac{1}{6}, -\frac{2}{6}, -\frac{3}{6}$ (\textit{cf.} Int), $-\frac{4}{6}, -\frac{5}{6}, -\frac{9}{10}$ (\textit{cf.} Low) from the baseline (\textit{cf.} High) value of the secretion rate of the chemoattractant; $+3, +6, +9$ (\textit{cf.} Int), $+12, +15, +18$ (\textit{cf.} Low) from the baseline (\textit{cf.} High) value of the TC-CTL adhesion strength; $+0.3, +0.15, -0.15, -0.3, -0.45, -0.5 $ (\textit{cf.} Low) from the baseline (\textit{cf.} Normal) value of the CTL growth rate. For each pair of parameters, the value of the third parameter is set to its baseline value. The parameter values used to carry out numerical simulations are provided in Tab.\ref{ch4:table1} and Tab.\ref{ch4:table2}.)
				}
				\label{ch4:heatmap}
			\end{figure}
			\subsection{Increasing the immune success rate has an impact on the infiltration score only when the TC-CTL adhesion strength is sufficiently large}
			\label{Increasing the immune success rate has an impact on the infiltration score only when the TC-CTL adhesion strength is high}
			So far, we have investigated with our model the effects of psychological stress on immune infiltration in the case where CTLs are not able to eliminate tumour cells.  However, in~\citep{al2021integrated} is reported that immune cells are activated against the spheroids, although the cytotoxic effect of immune cells on tumour cells is not particularly pronounced. 
			Motivated by these considerations, now we investigate tumour-immune dynamics and the effects of psychological stress on immune infiltration in the case where CTLs can eliminate tumour cells with a small probability.
			
			Figs.\ref{ch4:mainres3}\textbf{(a)}-\textbf{(b)} show a comparison between the infiltration score obtained with the parameter values considered in Sec.~\ref{Decreasing the secretion rate of the chemoattractant reduces the infiltration of CTLs into the tumour}, assuming that CTLs are able, or not able, to eliminate tumour cells (\textit{i.e} the immune success rate is either zero or different from zero - \textit{cf.} Tab.\ref{ch4:table2}).
			Figs.\ref{ch4:mainres3}\textbf{(c)}-\textbf{(d)} show a comparison between the number of tumour cells and CTLs at the end of simulations (corresponding to day 4 of the experiments) for the two scenarios considered. Comparing the results of Fig.\ref{ch4:mainres3}\textbf{(a)} with those displayed in Fig.\ref{ch4:mainres3}\textbf{(b)}, we see that, when  the TC-CTL adhesion strength is sufficiently high, increasing the immune success rate decreases the infiltration score. This is probably due to the fact that, 
			when CTLs can infiltrate through tumour cells, they are more likely to come into contact with tumour cells, thus increasing the chance for CTLs to eliminate them. Since dead tumour cells are cleared from the domain, this in turn diminishes the number of CTLs surrounded by tumour cells, leading to a reduced infiltration score. 
			However, when the TC-CTL adhesion strength is sufficiently low, increasing the immune success rate does not have an impact on the infiltration score. In fact, in this scenario, CTLs accumulate around the tumour, decreasing their probability to come into contact with tumour cells. This reduces their chance to eliminate tumour cells. Analogous considerations hold for the case in which lower growth rates of CTLs  are considered (results not shown).
			
			As shown by Figs.\ref{ch4:mainres3}\textbf{(c)}-\textbf{(d)}, increasing the immune success rate leads to a slightly decreased number of tumour cells at the end of simulations only when sufficiently large values of
			the TC-CTL adhesion strength and the secretion rate of the chemoattractant are considered. On the other hand, for intermediate and sufficiently small values of these two parameters, increasing the immune success rate does
			not have an impact on the final number of tumour cells. 
			
			\begin{figure}[!t]
				\centering
				\includegraphics[width=16.5cm]{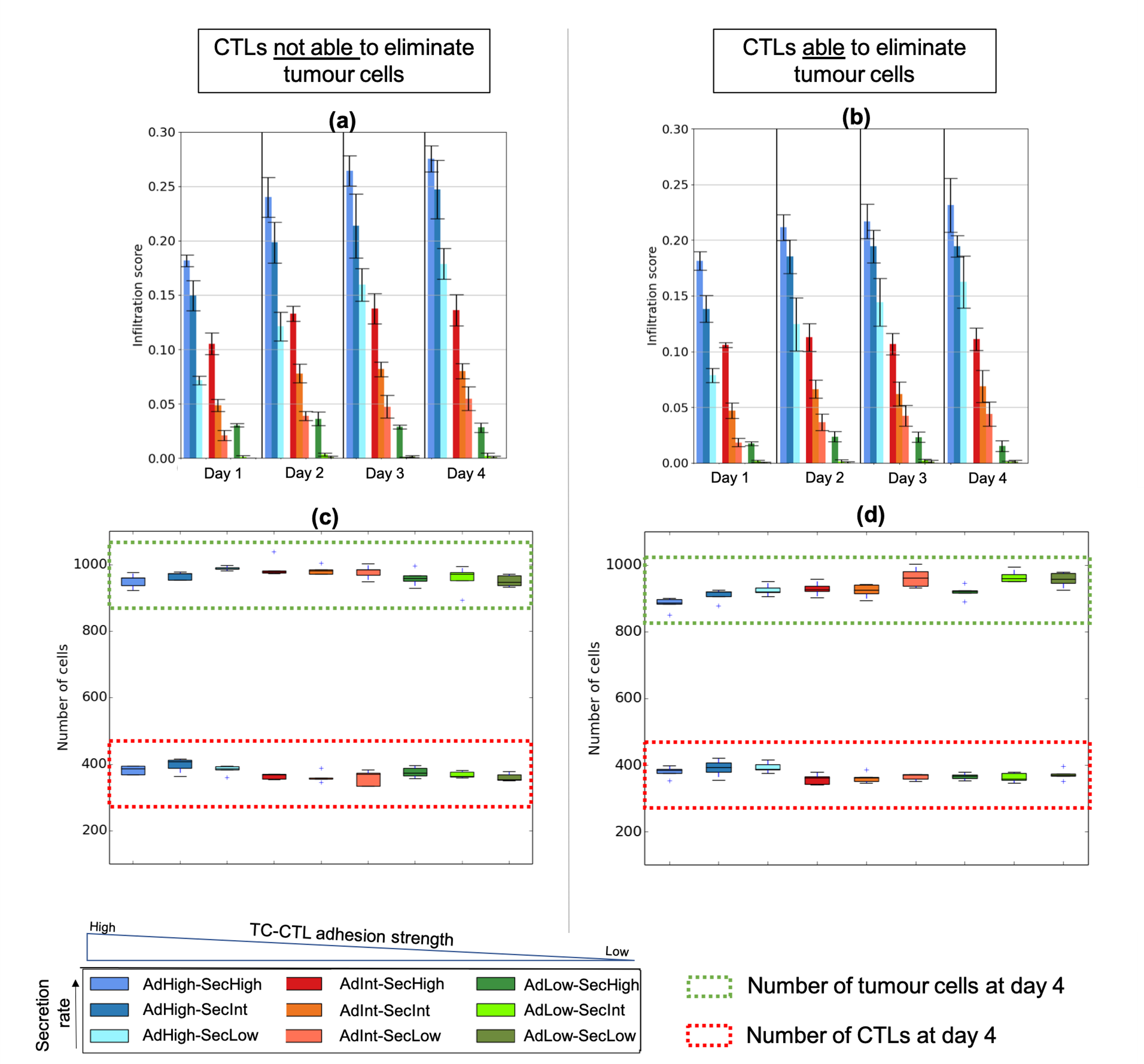}
				\caption{\textbf{Increasing the immune success rate has an impact on the infiltration
						score only when  the TC-CTL adhesion strength is sufficiently large.} Panel \textbf{(a)-(b)} displays the
					average value of the infiltration score, computed via \eqref{infiltration score}, for different values of the secretion rate of the chemoattractant and the TC-CTL adhesion strength, and at different times of the simulations. In panel \textbf{(a)} CTLs are assumed not to be able to eliminate tumour cells, while they are assumed to be able to eliminate them in panel \textbf{(b)}. Panels \textbf{(c)-(d)} display the
					corresponding number of tumour cells and CTLs at the end of simulations (corresponding to day
					4 of the experiments). The parameter values used to carry out numerical simulations are provided in Tab.\ref{ch4:table1} and Tab.\ref{ch4:table2}.
				}
				\label{ch4:mainres3}
			\end{figure}

			\section{Discussion, conclusions and research perspectives}
			\label{ch4:Discussion, conclusions and research perspectives}
			The \textit{in vitro} co-culture experiments presented in~\citep{al2021integrated} are performed in an isolated and relatively homogeneous environment and involve only a few constituents: tumour spheroids, activated immune cells, culture medium and cortisol. Furthermore, each experiment has clear observables, namely the confocal images of the co-culture, the trafficking indices  and the levels of IFN-$\gamma$ and IL-10, which make these experiments highly suitable to be studied through a mathematical model.

			In this paper, we have presented an IBM to describe the interaction dynamics between CTLs and tumour cells, to reproduce qualitative aspects presented in \citep{al2021integrated} and evaluate immune cell trafficking into tumour cells under normal and stressed conditions. 
			In particular, on the basis of the experiments presented in~\citep{al2021integrated}, we have investigated in a causal, systematic
			manner the way in which IFN-$\gamma$ and IL-10 may impact on the infiltration of CTLs into tumour cells. 
			
			The results of numerical simulations qualitatively reproduce, both in 2D and 3D,
			the growth of the tumour spheroids prior the introduction of immune cells and the tumour-immune dynamics in non-stressed conditions. The tumour growth is of logistic type. 
			In the control scenario, \textit{i.e.} the scenario in which the secretion rate of the chemoattractant, the TC-CTL adhesion strength and the CTLs growth rate are set at their baseline values, CTLs are able to infiltrate into the tumour.  
			
			We then have investigated the effects of psychological stress on immune infiltration. First, the results of our model support the idea that reducing the secretion rate of the chemoattractant and the
			TC-CTL adhesion strength, which are associated to a decrease in IFN-$\gamma$ levels, reduces the infiltration of CTLs into the tumour. These results also suggest that the secretion rate of the chemoattractant is more likely to have an impact on T cell infiltration when CTLs display a sufficiently high capability to infiltrate through tumour cells. 
			We have also studied the effects of psychological stress on immune infiltration by reducing the growth rate of CTLs, which is associated to increased IL-10 levels. Decreasing the growth rate of CTLs reduces the number of CTLs in the domain. This leads to a significant reduction in the infiltration score only when the TC-CTL adhesion strength is sufficiently large. The sensitivity analysis of these three paramaters has allowed us to undertake a more comprehensive investigation of the relationship between them and the value of the infiltration score.
			Finally, we have performed numerical simulations by letting CTLs eliminate tumour cells with a small probability - \textit{i.e.} when the immune success rate is greater than 0. In the scenario in which CTLs are able to infiltrate into the tumour, increasing the immune success rate leads to a reduced infiltration score, as tumour cells in contact with CTLs are eliminated. This in turn leads to a slightly decreased number of tumour cells at the end of simulations. 
			

			In summary, the results of numerical simulations of our model indicate that the interplay between IFN-$\gamma$ and IL-10 plays a key role in determining the effects of psychological stress on immune infiltration reported in \citep{al2021integrated}, as both cytokines contribute to regulate immune infiltration in opposite ways.
			Moreover, our results shed light on the impact of three biological stress-induced mechanisms on immune infiltration. In particular, they support the idea that a high infiltration score can be obtained only when the secretion rate of the chemoattractant and the TC-CTL adhesion strength are large, provided that the growth of CTLs is not inhibited. On the other hand, reducing the value of these parameters can lead to a reduced immune infiltration in different ways. For example, we found that the parameter having the strongest impact on immune infiltration is the TC-CTL adhesion strength, which is associated with the physical capability of CTLs to infiltrate through tumour cells.  In this regard, the development of abnormal structural features that inhibit the ability of CTLs to penetrate tumour sites is a hallmark of cancer progression~\citep{galon2019approaches}. Evidence is emerging that glucocorticoids act on adhesion of immune cells by inhibiting adhesion molecules (integrins and selectins) \citep{cronstein1992mechanism,kalfeist2022impact}. The deregulation of adhesion molecules may act as barriers to T cell migration and infiltration. In this context, the results of this study
			support the idea that new glucocorticoid receptor antagonists should be developed to target cell adhesion molecules in order to enhance immune infiltration.
			
			The results of numerical simulations also support the idea that an efficient anti-tumour immune response can occur only in highly infiltrated tumours. This is a key result because it indicates that therapeutic strategies promoting the infiltration of CTLs into tumours may be a promising approach against cancer. In particular, our findings suggest that a synergistic
			effect can be achieved by combining glucocorticoid receptor antagonists, which facilitate CTL infiltration, with immune checkpoint therapies, which enhance the
			effectiveness of \textit{in situ} anti-tumour immune response~\citep{galon2019approaches}. 
			
			The current version of our model can be developed further in several ways. Firstly, due to the high computational cost in simulating the three dimensional version of our model, we carried out 3D simulations only to part of our study. However, by running the simulations on high performance computers, this limitation may be addressed in the future and a larger spectrum of parameter values could be tested. 
			
			We managed to calibrate some parameters of the model (see Tab.\ref{ch4:table1} and Tab.\ref{ch4:table2}) from the literature and define them on the basis of precise biological assumptions. However, there are some parameters (\textit{e.g} parameters related to the dynamics of the chemoattractant, the death rate of tumour and CTLs due to intra-population competition) whose values were simply chosen with an exploratory aim and to qualitatively reproduce essential aspects of the experimental results obtained in~\citep{al2021integrated}. In order to minimise the impact of this limitation on the conclusions of our study, first we selected a baseline set of parameters that allowed to reproduce the growth of the spheroids and CTL infiltration as obtained in~\citep{al2021integrated}. Then, we carried out simulations by keeping all parameter fixed and changing only the values of our three parameters of interest, and then comparing the simulation results so obtained.
			
			To keep the model as simple as possible, we chose to include only mechanisms that were necessary to reproduce part of the experimental results presented in~\citep{al2021integrated}. If experimental measurements were available for cortisol, IFN-$\gamma$ and IL-10, we could calibrate the parameters related to their concentration dynamics, and then update the model in order to explicitly incorporate the dynamical modelling of these quantities  using PDEs.
			Also, our current model does not consider the effects of tumour necrosis and hypoxia or CTL exhaustion. These mechanisms can actively contribute to deregulate the normal levels of pro- and anti-inflammatory cytokines, resulting in more aggressive tumours and impaired immune response~\citep{galon2019approaches,balkwill2009tumour,jiang2015t,wherry2011t}. 
			
			From a biological point of view, a natural development of this work would consist in studying the effects of therapeutic strategies which counteract the negative impact of psychological stress on immune infiltration. In fact, in~\citep{al2021integrated} it was found that the administration of glucocorticoid receptor antagonists reversed the effects of cortisol and significantly enhanced immune infiltration in tumour spheroids. The effects of therapeutic strategies could be incorporated into our model by, for example, including a detailed metabolic network at the sub-cellular level that directly influences the dynamics at the cellular level, such as CTL growth and movement. 
			In this regard, we could also investigate the delivery schedule of therapeutic agents (\textit{i.e.} time and dosage) that may make it possible to maximise the number of infiltrated CTLs at the end of the treatment. 
			
			Despite its relative simplicity, our model provides a novel \textit{in silico} framework to investigate the impact of biological mechanisms linked to psychological stress on immune infiltration, and may be a promising tool to easily and cheaply explore therapeutic strategies designed to increase immune infiltration and improve the overall anti-tumour immune response. 
			\section*{Funding}
			E.L. has received funding from the European Research Council (ERC) under the European Union’s Horizon 2020 research and innovation programme (grant agreement No 740623). T.L. gratefully acknowledges support from the the PRIN 2020 project (No. 2020JLWP23) “Integrated Mathematical Approaches to Socio-Epidemiological Dynamics” (CUP: E15F21005420006). T.L. gratefully acknowledge support of the Institut Henri Poincar\'e (UAR 839 CNRS-Sorbonne Universit\'e), and LabEx CARMIN (ANR-10-LABX-59-01). L.A., T.L. and E.L. gratefully acknowledge support from the CNRS International Research Project ‘Mod\'elisation de la biom\'ecanique cellulaire et tissulaire’ (MOCETIBI).
	\newpage
	\begin{appendices}	
			\section{Details of the individual-based model}
		\label{Details of computational model}
		The individual-based model (IBM) has been numerically simulated using the multicellular modelling environment CompuCell3D~\citep{izaguirre2004compucell}. This software is an open source solver, which uses a Cellular Potts (CP) model~\citep{graner1992simulation} (also known as Glazier-Graner-Hogeweg model). In CP models, biological cells are treated as discrete entities represented as a set of lattice sites, defined as pixels in 2D (or voxels in 3D), each with characteristic values of area and perimeter (or volume and surface in 3D), and intrinsic motility on a regular lattice. Interaction descriptions and dynamics between cells are modelled by means of the effective energy of the system. This determines many characteristics such as cell size, motility, adhesion strength, and the reaction to gradients of chemotactic fields. During a simulation, each cell will attempt to extend its boundaries, through a series of index-copy attempts, in order to minimise the effective energy. The success of the index copy attempt depends on rules which take into account energy changes. 
		
		Files to run a simulation example of the model with Compucell3D \cite{izaguirre2004compucell} are available at \url{https://plmlab.math.cnrs.fr/leschiera/roleofstress}.
		
		\subsection{Cell types}
		In CP models, cells are uniquely identified with an index $\sigma_i$ on each lattice site $i$, with $i$ a vector of integers occupying lattice site $i$. Each cell in the model has a type $\tau(\sigma_i)$, which determines its properties, and the processes and interactions in which it participates. Note
		that, technically, the extracellular medium is also considered as a cell of type medium. In our model, we define 3 cell types: medium, tumour cell and CTL.
		\subsection{Cellular dynamics}
		The effective energy is the basis for operation of all CP models, including CompuCell3D~\citep{izaguirre2004compucell}, because it determines the interactions between cells (including the extracellular medium). Configurations evolve to minimise the effective energy $H$ of the system, defined in a two-dimensional system as
		
		\begin{equation}
			\footnotesize
			\begin{aligned}
				H&=\underbrace{\sum_{i,j}J(\tau(\sigma_i),\tau(\sigma_j))(1-\delta(\sigma_i,\sigma_j))}_{\text{boundary energy}}+\underbrace{\sum_\sigma\left[\lambda_{area}(\sigma)(a(\sigma)-A_t(\sigma))^2\right]}_{\text{area constraint}}+\underbrace{\sum_\sigma\left[\lambda_{per}(\sigma)(p(\sigma)-P_t(\sigma))^2\right]}_{\text{perimeter constraint}}.
				\label{Hsum}
			\end{aligned}
		\end{equation}
		The most important component of the effective energy equation is the boundary energy, which governs the adhesion of cells. The boundary energy $J(\tau(\sigma_i),\tau(\sigma_j))$ describes the contact energy between two cells $\sigma_i$ and $\sigma_j$ of types $\tau(\sigma_i)$ and $\tau(\sigma_j)$. It is calculated by summing over all neighbouring pixels $i$ and $j$ that form the boundary between two cells. Moreover, $\delta(\sigma_i,\sigma_j)=1$ if $\sigma_i=\sigma_j$, and $\delta(\sigma_i,\sigma_j)=0$ otherwise. Thanks to the term $(1-\delta(\sigma_i,\sigma_j))$, the boundary energy contribution is considered only between lattice sites belonging to two different cells. When considering a two-dimensional domain, the second and third terms represent, respectively, a cell-area and cell-perimeter constraint. In particular, $a(\sigma)$ and $p(\sigma)$ are the surface area and perimeter of the cell $\sigma$, $A_t(\sigma)$ and $P_t(\sigma)$ are the cell’s target surface area and perimeter, and $\lambda_{area}(\sigma)$ and $\lambda_{per}(\sigma)$ are an area and perimeter constraint coefficients. Note that in 3D these two terms  represent, respectively, a cell-volume and cell-surface constraint and, therefore, they might assume a different value. 
		
		The cell configuration evolves through lattice-site copy attempts. To begin an index-copy attempt, the algorithm randomly selects a lattice site to be a target pixel $i$, and a neighbouring lattice site to be a source pixel $i^\prime$. If the source and target pixels belong to the same cell (\textit{i.e.} if $\sigma_i =\sigma_{i^\prime}$, they do not need to attempt an lattice-site copy and thus the effective energy will not be calculated. Otherwise, an attempt will be made to switch the target pixel as the source pixel, thereby increasing the surface area of the source cell and decreasing the surface area of the target cell.  \\
		The algorithm computes $\Delta H=H-H^\prime$, with $H$ being the effective energy of the system and $H^\prime$ being the effective energy if the copy occurs. Then, it sets $\sigma_i =\sigma_{i^\prime}$ with probability $P(\sigma_i\rightarrow\sigma_{i^\prime})$ given by
		\begin{equation}
			P(\sigma_i\rightarrow\sigma_{i^\prime})= \begin{cases}\quad 1 \; \;  \quad : \quad \Delta H \le 0 \\
				\exp^{-\frac{\Delta H}{T_m}} \; : \quad \Delta H > 0.
			\end{cases}
			\label{Boltzmann}
		\end{equation}
		The change in effective energy $\Delta H$ provides a measure of the energy cost of such a copy and the parameter~$T_m$ determines the level of stochasticity of accepted copy attempts. The
		unit of simulation time is the Monte Carlo step (MCS).
		\subsection{Subcellular dynamics and chemotaxis}
		In our model we simulate CTL chemotaxis toward tumour cells, defined as the cell motion induced by the presence of a chemical gradient. In CompuCell3D~\citep{izaguirre2004compucell}, chemotaxis is obtained biasing the cell’s motion up or down a field gradient by adding a term $\Delta H_{chem}$ in the calculated effective-energy change $\Delta H $ used in the acceptance function \eqref{Boltzmann}. For a field $c_i$: 
		\begin{equation}
			\Delta H_{chem}=-\lambda_{chem}(\phi_i-\phi_{i^\prime})),
			\label{chem}
		\end{equation}
		where $\phi_i$ is the chemical field at the index-copy target pixel $i$, $\phi_{i^\prime}$ the field at the index-copy source pixel $i^\prime$, and $\lambda_{chem}\geq 0$ the strength of chemotaxis. \\
		The change in concentration of the chemical field $\phi$ is obtained by solving a reaction-diffusion equation of the following general form: 
		\begin{equation}
			\frac{\partial \phi}{\partial t}= D \Delta\phi-\gamma\phi+S  
		\end{equation}
		where $D$, $\gamma$ and $S$ denote the diffusion constant, decay constant and secretion rates of the field, respectively. These three parameters may vary with position and cell-lattice configuration, and thus be a function of cell~$\sigma$ and pixel $i$.
		
		In CompuCell3D, this general form of PDEs can be solved using a number of different PDE solvers. More details about the different PDE solvers can be found in the CompuCell3D Reference Manual.
		
		In the main body of the paper, the dynamic of the concentration of the chemoattractant secreted by tumour cells $\phi$ is governed by the following reaction-diffusion equation:
		\begin{equation}
			\frac{\partial \phi}{\partial t}= D \Delta \phi- \gamma \phi+ \alpha\sum_{n=1}^{N_T(t)}\delta_{{n\in N_{TB}(t)}}.
			\label{inf gamma}
		\end{equation}
		In Eq.~\eqref{inf gamma}, $D$ is the chemoattractant diffusion constant, $\gamma$ is the rate of natural decay and $\alpha$ is the secretion rate. Moreover, $\delta_{{n\in N_{TB}(t)}}=1$ if $n\in N_{TB}(t)$, and $\delta_{{n\in N_{TB}(t)}}=0$ otherwise, where $N_{TB}(t)$ denotes the set of indices of tumour cells in contact with the surrounding medium at time $t$. This terms, handled in Compucell3D by the DiffusionSolverFE, takes into account the fact that only the tumour cells at the border of the tumour secrete the chemoattractant. We complement Eq.~\eqref{inf gamma} with zero-flux boundary conditions and an initial concentration $\phi_{init}$ at time 0 of the experiments (\textit{i.e.} when CTLs are introduced) which is set to be zero
		everywhere in the domain but at the border of the tumour (\textit{cf.} Tab.\ref{ch4:table2}).
		\subsection{Infiltration score}
		Building on the TIC proposed in \cite{al2021integrated} to measure immune infiltration levels, in our model at each time-step we compute the ‘infiltration score’ via Eq.~\eqref{infiltration score}. This score allows us to quantify the level of CTL infiltration
		into the tumour and is defined as the number of CTLs surrounded by tumour cells, divided by the number
		of tumour cells and CTLs surrounded by tumour cells. Below we detail how the infiltration score is implemented in CompuCell3D.   
		
		In  CompuCell3D, the function $get\_cell\_neighbor\_data\_list(cell)$ allows to access a list of each cell neighbors. The neighbour of a cell is defined as an adjacent cell that shares a surface area with the cell in question. 
		In our model, for each cell we loop over all its neighbors and we compute its common surface area with medium, tumour cells and CTLs using the function $neighbor\_list.common\_surface\_area\_by\_type(cell)$. We then assume that $\delta_{{m\in N_{CS}(t)}}=1$ if the common surface area between the $m^{th}$ CTL and CTLs and medium surrounding it is strictly lower than 4 and that $\delta_{{n\in N_{TS}(t)}}=1$ if the common surface area between the $n^{th}$ tumour cell and CTLs and medium surrounding it is strictly lower than 6. More details of these functions can be found in the CompuCell3D manuals available at: \url{https://compucell3d.org/Manuals}.
		\section{Initial conditions and values of model parameters}
		\label{appendix5:graph}
		
		The IBM is based on the mathematical model developed in our previous work~\cite{leschiera2021mathematical}, and has been calibrated to qualitatively reproduce the experimental results presented in~\cite{al2021integrated}. 
		
		\subsection{Set-up of simulations}
		To carry out numerical simulations in Sec.~\ref{ch4:Numerical simulations and preliminary results}, we used a CP approach both on a 2D spatial domain with a total of $400\times 400$ lattice sites and on a 3D spatial domain with a total of $100\times 100 \times 100$ lattice sites. The numerical simulations we present in Sec.~\ref{ch4:Main results} were carried out on the 2D domain only. In both cases, simulations were performed using the software CompuCell3D~\cite{izaguirre2004compucell} on a standard workstation (Intel i7 Processor, 4 cores, 16 GB RAM, macOS 11.2.2). 
		
		
		At the initial time point of simulations (\textit{i.e.} on day -7), 35 tumour cells are placed in the centre of the domain (\textit{cf.} Fig.\ref{ch4:tumour_no_T_cells}). First we let tumour cells grow in the absence of CTLs for 11 days, carrying out numerical simulations for $33000$ time-steps. On day 0, the number of tumour cells is set at carrying capacity (\textit{i.e.} 950 cells). This is done to ensure that the changes in the tumour surface and volume are due to the infiltration of CTLs into the tumour. We then randomly introduce 150 CTLs at the border of the domain and we let them grow, move and interact with tumour cells for 4 days, carrying out numerical simulations for $1200$ time-steps (\textit{cf.} Figs.\ref{ch4:baselineTOT}-\ref{ch4:mainres3}). 
		
		In the next subsection we describe the way in which additional components of the model were calibrated leading to the parameter values reported in Tab.\ref{ch4:table1} and Tab.\ref{ch4:table2}, to qualitatively reproduce the behaviour of the experimental results presented in~\cite{al2021integrated}. 
		
		\subsection{Model calibration through parameter exploration} The model is calibrated to qualitatively reproduce the experimental results reported in~\cite{al2021integrated}. 
		Due to computational cost, it was not feasible to start with the actual number of cells present within a cell culture (which can reach the order of magnitude $10^5$) or to simulate the same number of  cells as found in a real tumour. With our model, we instead focused on qualitatively capturing the change in immune infiltration levels while varying a certain set of parameters. 
		
		Some parameters of the model (see Tab.\ref{ch4:table1} and Tab.\ref{ch4:table2}) are estimated from the literature
		and defined on the basis of precise biological assumptions. Other model parameters that could not be based on a literature source, such as the cell death rates due to intra-population competition, are adjusted to qualitatively reproduce the growth of the spheroids and CTLs in non-stressed conditions presented in \cite{al2021integrated}. Finally, there are some parameters, such as the TC-CTL adhesion strength, the immune success
		rate and the secretion rate of the chemoattractant, whose values were simply chosen and varied with an exploratory
		aim to qualitatively reproduce essential aspects of the experimental results obtained in~\cite{al2021integrated}.
		
		\subsection{Parameters in the 2D system}
		In the 2D system, the value of the rate of death due to competition
		between tumour cells is chosen so that the number of tumour cells reaches its carrying capacity after 7 days of proliferation. The number of CTLs introduced in the domain on day 0 and the value of the rate of death due to competition
		between CTLs are chosen so that the value of the infiltration score computed at the end of the 2D simulations in the control scenario is similar to the value of the TIC algorithm obtained on day 4 in~\cite{al2021integrated}, when cortisol is not introduced in the
		co-culture. When modelling tumour cell and CTL growth, at each time-step, we let cells grow at a random rate drawn from a uniform distribution; the parameters of the bounds of the uniform distribution are chosen to match the mean duration of a tumour cell and CTL cycle length.
		
		The ratio between the energy at the interface between tumour cells and CTLs and the energy at the interface between tumour cells (\textit{i.e.} the values of parameters $J_{CT}$ and $J_{TT}$ in Eq. \eqref{Hsum}) allows us to consider a wide range
		of biological scenarios corresponding to different degrees of immune infiltration. In particular, if $J_{CT} < J_{TT}$, then CTLs infiltrate through tumour cells, whereas if $J_{CT} > J_{TT}$, then CTLs accumulate at the margin of the tumour, without infiltrating it.  Therefore, to obtain different degrees of immune infiltration, we fix the value of $J_{TT}$ and we vary the value of $J_{CT}$. In the body of the paper we refer to the parameter $J_{CT}$ as “tumour cell-CTL adhesion strength” (TC-CTL adhesion strength). In the control scenario we suppose that CTLs have a high capability of infiltrate through tumour cells. Therefore, we suppose that $J_{CT} < J_{TT}$. In stressed conditions instead, we suppose that CTLs have a lower capability to infiltrate through tumour cells. Therefore, we increase the value of $J_{CT}$ to a value equal to or greater than that chosen for  $J_{TT}$.
		
		\subsection{Parameters in the 3D system}
		The numerical simulations shown in Sec.~\ref{ch4:Numerical simulations and preliminary results} attempted to verify that our model produces similar results
		both in 2D and 3D. Therefore, in the 3D system we make use of the same values selected in the two-dimensional case (see previous subsection). Due to the slightly different number of cells obtained at the end of the numerical simulations in 3D, we simply adjust the death rate of tumour cells and CTLs due to intra-population competition, in order to obtain a number of cells at the end of the numerical simulations similar to that of the 2D scenario.
		\begin{table}[hp!]
			\small
			\caption{Parameter values used to implement the CP model. Energies, temperature and constrains are dimensionless parameters.}
			\centering
			\begin{tabular}{p{1.3cm} p{1.2cm} p{6.3cm} p{4.5cm} p{1cm}}
				\hline
				Phenotype & Symbol & Description & Value & Ref. \\
				\hline
				\textbf{Domain} & Pixel & Lattice site in 2D & $1$ Pixel $=3\times3\; \mu m^2$ &  \\ 
				& Voxel & Lattice site in 3D & $1$ Voxel $=3\times3\times3\; \mu m^3$ &  \\ 
				& $\Delta t$ &Time-step&  1 MCS = 0.5 min & \\
				\\
				\textbf{CC3D} & $J_{MT}$ & Contact energy tumour cells-medium & 50 &  ~\cite{leschiera2021mathematical}\\ 
				& $J_{MC}$ & Contact energy CTLs-medium & 50 & ~\cite{leschiera2021mathematical} \\ 
				& $J_{CT}$ & Contact energy CTLs-tumour cells & high adh.: 5; intermediate adh.: 50; low adh.: 95  &  \\ 
				& $J_{TT}$ & Contact energy tumour cells-tumour cells & 50 & ~\cite{leschiera2021mathematical} \\ 
				& $J_{CC}$ & Contact energy CTLs-CTLs & 1000 & ~\cite{leschiera2021mathematical} \\
				& $d_{T}$ & Tumour cell diameter  & 20-40 $(\mu m)$ &~\cite{gong2017computational}\\
				& $d_{C}$ & CTL diameter & 12 $(\mu m)$ & ~\cite{gao20162}\\
				& $A_0$ & Initial area constrain (2D) & $\mathcal{U}_{[25,55]}$ - tumour cells \textit{(pixels)} $\mathcal{U}_{[20,25]}$ - CTLs \textit{(pixels)} &  \\
				& $V_0$ & Initial volume constrain (3D) & $\mathcal{U}_{[25,55]}$ - tumour cells  \textit{(voxels)} $\mathcal{U}_{[20,25]}$ - CTLs  \textit{(voxels)} &  \\
				& $P_t$ & Perimeter constrain (2D) & $4\sqrt{A_t}+0.5$ \textit{(pixels)} &  \\
				& $S_t$ & Surface constrain (3D) & $6V_t^{\frac{2}{3}}$ \textit{(voxels)} &  \\
				& $\lambda_{area}$ & Tumour cell and CTL area constrain (2D)  & $10$ & \\
				& $\lambda_{per}$ & Tumour cell and CTL perimeter constrain (2D)  & $10$ & \\
				& $\lambda_{vol}$ & Tumour cell and CTL volume constrain (3D)  & $20$ & \\
				& $\lambda_{surf}$ & Tumour cell and CTL surface constrain (3D)  & $20$ & \\
				& $T_m$ & Fluctuation amplitude parameter & $10$ & ~\cite{leschiera2021mathematical}\\
				& $\lambda_{chem}$ & Strength and direction of chemotaxis  & $50$ & ~\cite{leschiera2021mathematical}\\
				\hline
			\end{tabular}
			\label{ch4:table1}
		\end{table}
		
		\begin{table}[hp!]
			\small
			\caption{Parameter values used in numerical simulations.}
			\centering
			\begin{adjustbox}{angle=0}
				\begin{tabular}{p{1.8cm} p{4.3cm} p{6.6cm} p{1cm}}
					\hline
					Phenotype  & Description & Value & Ref. \\
					\hline
					\textbf{Tumour} & Initial number  &$N_T(0) =36$ &  \\ 
					& Index identifier& $n=1, \dots, N_T(t)$&  \\
					
					&  Lifespan   & $\mathcal{U}_{[3,7]}$ (\textit{days}) & ~\cite{gong2017computational}\\
					
					& Growth rate  &  $\mathcal{U}_{[0.015,0.019]}$ (\textit{pixel or voxel/MCS})  &  ~\cite{al2021integrated}
					\\
					
					& Mean cycle time  & $12$ (\textit{hours}) & ~\cite{al2021integrated}  
					\\
					& Rate of death due to intra-pop. competition   & (2D) $4.6\times 10^{-7}$ (\textit{1/MCS}) {\color{white}-- - - - - - - - - - - - } (3D) $4.4\times10^{-7}$ (\textit{1/MCS}) &  estim. estim. 
					\\\\
					\textbf{CTLs}  &Initial number & $N_C(0)=150 $&  \\
					& Index identifier  & $m=1, \dots, N_C(t)$&  \\
					& Growth rate  & normal: $\mathcal{U}_{[0.0038,0.0042]}$ (\textit{pixel or voxel/MCS})   decreased:  $\frac{1}{2}\mathcal{U}_{[0.0038,0.0042]}$ (\textit{pixel/MCS})& ~\cite{al2021integrated}
					\\
					& Mean cycle time  & 8-10 (\textit{hours}) & ~\cite{al2021integrated,gong2017computational}  
					\\
					& Rate of death due to intra-pop. competition   & (2D) $1.2\times 10^{-6}$ (\textit{1/MCS})  {\color{white}-- -  - - - - - - - - - -} (3D) $1.3\times 10^{-6}$ (\textit{1/MCS}) &  estim.  estim.
					
					\\
					& Lifespan & $\mathcal{U}_{[2.5,3.5]}$ (\textit{days})&  ~\cite{gong2017computational}\\
					& Engagement time & 6 (\textit{hours})  & ~\cite{christophe2015biased} \\
					& Immune success rate  & Figs~\ref{ch4:tumour_no_T_cells}-\ref{ch4:mainres2}:	$0$; Fig~\ref{ch4:mainres3}: $0.00005$ &   \\
					\\
					\textbf{Chemoattr.} 
					& Concentration &  $\phi\geq 0 $ (\textit{mol/pixel or voxel}) &  \\
					& Diffusion & $ D=2$ (\textit{pixel$^2$ or voxel$^3$/MCS}) &  ~\cite{leschiera2021mathematical}\\
					& Secretion &  high: $\alpha=30$; intermediate: $\alpha=10$; low: $\alpha=3$  (\textit{mol/MCS/pixel or voxel})\\
					& Decay & $\gamma=7\times 10^{-4} $ (\textit{1/MCS}) & ~\cite{leschiera2021mathematical} \\
					& Initial concentration  & (2D) $\phi^{init}=0.5(280-\sqrt{(x-200)^2+(y-200)^2})$  (3D) $\phi^{init}=0.5(280-\sqrt{(x-50)^2+(y-50)^2 +(z-50)^2})$&  \\
					
					\hline
				\end{tabular}
			\end{adjustbox}
			\label{ch4:table2}
		\end{table}
	\end{appendices}	 
	\bibliographystyle{vancouver}
	\bibliography{mainDocumentNoComment.bib}

\begin{thebibliography}{10}

\bibitem{coe2007psychosocial}
Coe CL, Laudenslager ML.
\newblock Psychosocial influences on immunity, including effects on immune
  maturation and senescence.
\newblock Brain, behavior, and immunity. 2007;21(8):1000-8.

\bibitem{morey2015current}
Morey JN, Boggero IA, Scott AB, Segerstrom SC.
\newblock Current directions in stress and human immune function.
\newblock Curr Opin Psychol. 2015;5:13-7.

\bibitem{seiler2020impact}
Seiler A, Fagundes CP, Christian LM.
\newblock The impact of everyday stressors on the immune system and health.
\newblock Stress challenges and immunity in space: From mechanisms to
  monitoring and preventive strategies. 2020:71-92.

\bibitem{lee2009surgical}
Lee JW, Shahzad MMK, Lin YG, Armaiz-Pena G, Mangala LS, Han HD, et~al.
\newblock Surgical stress promotes tumor growth in ovarian carcinoma.
\newblock Clin Cancer Res. 2009;15(8):2695-702.

\bibitem{nilsson2007stress}
Nilsson MB, Armaiz-Pena G, Takahashi R, Lin YG, Trevino J, Li Y, et~al.
\newblock Stress hormones regulate interleukin-6 expression by human ovarian
  carcinoma cells through a {Src}-dependent mechanism.
\newblock J Biol Chem. 2007;282(41):29919-26.

\bibitem{budiu2017restraint}
Budiu RA, Vlad AM, Nazario L, Bathula C, Cooper KL, Edmed J, et~al.
\newblock Restraint and social isolation stressors differentially regulate
  adaptive immunity and tumor angiogenesis in a breast cancer mouse model.
\newblock J Clin Oncol. 2017;6(1):12.

\bibitem{al2021integrated}
Al-Hity G, Yang F, Campillo-Funollet E, Greenstein AE, Hunt H, Mampay M, et~al.
\newblock An integrated framework for quantifying immune-tumour interactions in
  a {3D} co-culture model.
\newblock Commun Biol. 2021;4(1):1-12.

\bibitem{dranoff2004cytokines}
Dranoff G.
\newblock Cytokines in cancer pathogenesis and cancer therapy.
\newblock Nat Rev Cancer. 2004;4(1):11-22.

\bibitem{castro2018interferon}
Castro F, Cardoso AP, Gon{\c{c}}alves RM, Serre K, Oliveira MJ.
\newblock Interferon-gamma at the crossroads of tumor immune surveillance or
  evasion.
\newblock Front Immunol. 2018;9:847.

\bibitem{schiltz2002effects}
Schiltz PM, Gomez GG, Read SB, Kulprathipanja NV, Kruse CA.
\newblock {Effects of IFN-$\gamma$ and interleukin-1 $\beta$ on major
  histocompatibility complex antigen and intercellular adhesion molecule-1
  expression by 9L gliosarcoma: relevance to its cytolysis by alloreactive
  cytotoxic T lymphocytes}.
\newblock J Interferon Cytokine Res. 2002;22(12):1209-16.

\bibitem{alhakeem2018chronic}
Alhakeem SS, McKenna MK, Oben KZ, Noothi SK, Rivas JR, Hildebrandt GC, et~al.
\newblock Chronic lymphocytic Leukemia--Derived {IL-10 } suppresses antitumor
  immunity.
\newblock J Immun. 2018;200(12):4180-9.

\bibitem{couper200810}
Couper KN, Blount DG, Riley EM.
\newblock IL-10: the master regulator of immunity to infection.
\newblock J Immun. 2008;180(9):5771-7.

\bibitem{de2003mathematical}
de~Pillis LG, Radunskaya A.
\newblock A mathematical model of immune response to tumor invasion.
\newblock In: Computational fluid and solid mechanics 2003. Elsevier; 2003. p.
  1661-8.

\bibitem{kirschner1998modeling}
Kirschner D, Panetta JC.
\newblock Modeling immunotherapy of the tumor--immune interaction.
\newblock J Math Biol. 1998;37(3):235-52.

\bibitem{kuznetsov1994nonlinear}
Kuznetsov VA, Makalkin IA, Taylor MA, Perelson AS.
\newblock Nonlinear dynamics of immunogenic tumors: parameter estimation and
  global bifurcation analysis.
\newblock Bull Math Biol. 1994;56(2):295-321.

\bibitem{luksza2017neoantigen}
{\L}uksza M, Riaz N, Makarov V, Balachandran VP, Hellmann MD, Solovyov A,
  et~al.
\newblock A neoantigen fitness model predicts tumour response to checkpoint
  blockade immunotherapy.
\newblock Nature. 2017;551(7681):517-20.

\bibitem{delitala2013recognition}
Delitala M, Lorenzi T.
\newblock Recognition and learning in a mathematical model for immune response
  against cancer.
\newblock Discrete Contin Dyn Syst Ser B. 2013;18(4):891.

\bibitem{kolev2013numerical}
Kolev M, Nawrocki S, Zubik-Kowal B.
\newblock Numerical simulations for tumor and cellular immune system
  interactions in lung cancer treatment.
\newblock Commun Nonlinear Sci Numer Simul. 2013;18(6):1473-80.

\bibitem{lorenzi2015mathematical}
Lorenzi T, Chisholm RH, Melensi M, Lorz A, Delitala M.
\newblock Mathematical model reveals how regulating the three phases of
  {T-cell} response could counteract immune evasion.
\newblock Immunology. 2015;146(2):271-80.

\bibitem{atsou2020size}
Atsou K, Anju{\`e}re F, Braud VM, Goudon T.
\newblock A size and space structured model describing interactions of tumor
  cells with immune cells reveals cancer persistent equilibrium states in
  tumorigenesis.
\newblock J Theor Biol. 2020;490:110163.

\bibitem{atsou2021size}
Atsou K, Anju{\`e}re F, Braud VM, Goudon T.
\newblock A size and space structured model of tumor growth describes a key
  role for protumor immune cells in breaking equilibrium states in
  tumorigenesis.
\newblock PloS one. 2021;16(11):e0259291.

\bibitem{matzavinos2004travelling}
Matzavinos A, Chaplain MA.
\newblock Travelling-wave analysis of a model of the immune response to cancer.
\newblock C R Biol. 2004;327(11):995-1008.

\bibitem{matzavinos2004mathematical}
Matzavinos A, Chaplain MA, Kuznetsov VA.
\newblock Mathematical modelling of the spatio-temporal response of cytotoxic
  {T-lymphocytes to a solid tumour}.
\newblock Math Med Biol. 2004;21(1):1-34.

\bibitem{bouchnita2017hybrid}
Bouchnita A, Belmaati FE, Aboulaich R, Koury MJ, Volpert V.
\newblock A hybrid computation model to describe the progression of multiple
  myeloma and its intra-clonal heterogeneity.
\newblock Computation. 2017;5(1):16.

\bibitem{ghaffarizadeh2018physicell}
Ghaffarizadeh A, Heiland R, Friedman SH, Mumenthaler SM, Macklin P.
\newblock {PhysiCell: an open source physics-based cell simulator for 3-D
  multicellular systems}.
\newblock PLoS Comp Biol. 2018;14(2):e1005991.

\bibitem{almeida2022hybrid}
Almeida L, Audebert C, Leschiera E, Lorenzi T.
\newblock A Hybrid Discrete--Continuum Modelling Approach to Explore the Impact
  of T-Cell Infiltration on Anti-tumour Immune Response.
\newblock Bull Math Biol. 2022;84(12):1-37.

\bibitem{leschiera2021mathematical}
Leschiera E, Lorenzi T, Shen S, Almeida L, Audebert C.
\newblock {A mathematical model to study the impact of intra-tumour
  heterogeneity on anti-tumour CD8+ T cell immune response}.
\newblock J Theor Biol. 2022:111028.

\bibitem{christophe2015biased}
Christophe C, M{\"u}ller S, Rodrigues M, Petit AE, Cattiaux P, Dupr{\'e} L,
  et~al.
\newblock {A biased competition theory of cytotoxic T lymphocyte interaction
  with tumor nodules}.
\newblock PLoS ONE. 2015;10(3):e0120053.

\bibitem{kather2017silico}
Kather JN, Poleszczuk J, Suarez-Carmona M, Krisam J, Charoentong P, Valous NA,
  et~al.
\newblock {\textit{In silico}} modeling of immunotherapy and stroma-targeting
  therapies in human colorectal cancer.
\newblock Cancer Res. 2017;77(22):6442-52.

\bibitem{macfarlane2018modelling}
Macfarlane FR, Lorenzi T, Chaplain MA.
\newblock Modelling the immune response to cancer: an individual-based approach
  accounting for the difference in movement between inactive and activated {T}
  cells.
\newblock Bull Math Biol. 2018;80(6):1539-62.

\bibitem{macfarlane2019stochastic}
Macfarlane FR, Chaplain MA, Lorenzi T.
\newblock A stochastic individual-based model to explore the role of spatial
  interactions and antigen recognition in the immune response against solid
  tumours.
\newblock J Theor Biol. 2019;480:43-55.

\bibitem{izaguirre2004compucell}
Izaguirre JA, Chaturvedi R, Huang C, Cickovski T, Coffland J, Thomas G, et~al.
\newblock {CompuCell}, a multi-model framework for simulation of morphogenesis.
\newblock Bioinformatics. 2004;20(7):1129-37.

\bibitem{boissonnas2007vivo}
Boissonnas A, Fetler L, Zeelenberg IS, Hugues S, Amigorena S.
\newblock In vivo imaging of cytotoxic T cell infiltration and elimination of a
  solid tumor.
\newblock J Exp Med. 2007;204(2):345-56.

\bibitem{galon2019approaches}
Galon J, Bruni D.
\newblock Approaches to treat immune hot, altered and cold tumours with
  combination immunotherapies.
\newblock Nat Rev Drug Discov. 2019;18(3):197-218.

\bibitem{gorbachev2007cxc}
Gorbachev AV, Kobayashi H, Kudo D, Tannenbaum CS, Finke JH, Shu S, et~al.
\newblock CXC chemokine ligand 9/monokine induced by IFN-$\gamma$ production by
  tumor cells is critical for T cell-mediated suppression of cutaneous tumors.
\newblock J Immunol. 2007;178(4):2278-86.

\bibitem{tokunaga2018cxcl9}
Tokunaga R, Zhang W, Naseem M, Puccini A, Berger MD, Soni S, et~al.
\newblock CXCL9, CXCL10, CXCL11/CXCR3 axis for immune activation--a target for
  novel cancer therapy.
\newblock Cancer Treat Rev. 2018;63:40-7.

\bibitem{harjunpaa2019cell}
Harjunp{\"a}{\"a} H, Llort~Asens M, Guenther C, Fagerholm SC.
\newblock Cell adhesion molecules and their roles and regulation in the immune
  and tumor microenvironment.
\newblock Front Immunol. 2019;10:1078.

\bibitem{jorgovanovic2020roles}
Jorgovanovic D, Song M, Wang L, Zhang Y.
\newblock {Roles of IFN-$\gamma$ in tumor progression and regression: A
  review}.
\newblock Biomark Res. 2020;8(1):1-16.

\bibitem{cronstein1992mechanism}
Cronstein BN, Kimmel SC, Levin RI, Martiniuk F, Weissmann G.
\newblock A mechanism for the antiinflammatory effects of corticosteroids: the
  glucocorticoid receptor regulates leukocyte adhesion to endothelial cells and
  expression of endothelial-leukocyte adhesion molecule 1 and intercellular
  adhesion molecule 1.
\newblock Proc Natl Acad Sci. 1992;89(21):9991-5.

\bibitem{kalfeist2022impact}
Kalfeist L, Galland L, Ledys F, Ghiringhelli F, Limagne E, Ladoire S.
\newblock Impact of glucocorticoid use in oncology in the immunotherapy era.
\newblock Cells. 2022;11(5):770.

\bibitem{balkwill2009tumour}
Balkwill F.
\newblock Tumour necrosis factor and cancer.
\newblock Nat Rev Cancer. 2009;9(5):361-71.

\bibitem{jiang2015t}
Jiang Y, Li Y, Zhu B.
\newblock {T}-cell exhaustion in the tumor microenvironment.
\newblock Cell Death Dis. 2015;6(6):e1792-2.

\bibitem{wherry2011t}
Wherry EJ.
\newblock {T} cell exhaustion.
\newblock Nat Immunol. 2011;12(6):492-9.

\end{thebibliography}
		\end{document}